\documentclass[12pt]{article}
\usepackage{epsfig}
\usepackage{color}
\usepackage{amssymb,amsmath}
\usepackage{graphicx}

\newcommand{\bea}{\begin{eqnarray}}
\newcommand{\eea}{\end{eqnarray}}

 \makeatletter
\def\alt{\mathrel{\mathpalette\gl@align<}}
\def\agt{\mathrel{\mathpalette\gl@align>}}
\def\gl@align#1#2{\lower.6ex\vbox{\baselineskip\z@skip\lineskip\z@
\ialign{$\m@th#1\hfil##\hfil$\crcr#2\crcr\sim\crcr}}} \makeatother

\begin{document}
\begin{flushright}
\end{flushright}
\vspace*{1.0cm}

\begin{center}
\baselineskip 20pt 
{\Large\bf 
Gaugino Mediation Combined with the Bulk Matter Randall-Sundrum Model
}
\vspace{1cm}

{\large 
Nobuchika Okada$^{a}$ and Toshifumi Yamada$^{b,c}$
} \vspace{.5cm}

{\baselineskip 20pt \it
$^{a}$ Department of Physics and Astronomy, University of Alabama, 
Tuscaloosa, AL 35487, USA \\

$^{b}$ Department of Particles and Nuclear Physics, \\
The Graduate University for Advanced Studies (SOKENDAI) \\

$^{c}$ Institute of Particle and Nuclear Studies, \\ 
High Energy Accelerator Research Organization (KEK),  \\
1-1 Oho, Tsukuba, Ibaraki 305-0801, Japan} 

\vspace{.5cm}

\vspace{1.5cm} {\bf Abstract} \end{center}

We investigate a simple  5D extension of 
 the Minimal Supersymmetric (SUSY) Standard Model (SM) 
 that is combined with the bulk matter Randall-Sundrum (RS) model,
 which gives a natural explanation the Yukawa coupling hierarchy.
In this model, matter and gauge superfields reside in the 5D bulk 
 while a SUSY breaking sector and the Higgs doublet superfields
 are localized on the infrared brane. 
The Yukawa coupling hierarchy in SM 
 can be naturally explained through the wavefunction 
 localization of the matter superfields. 
While sparticles obtain their flavor-blind soft SUSY breaking masses 
 dominantly from the gaugino-mediated SUSY breaking, 
 flavor-violating soft terms arise through the gravity-mediated 
 SUSY breaking which are controlled by the wavefunction 
 localization of the matter superfields. 
This structure of the model allows us to predict 
 the sparticle mass spectrum including flavor-violating terms. 
We first explicitly determine the 5D disposition of matter superfields 
 from the low energy experimental data on SM fermion masses, 
 CKM matrix and the neutrino oscillation parameters. 
Then, we calculate particle mass spectra and estimate the effects 
 of the flavor-violating soft terms, which should be compared 
 with the current experimental constraints. 
With gravitino being the lightest sparticle (LSP), 
 the next-to-LSP, which is long-lived, is predicted 
 most likely to be either singlet smuon-like or selectron-like. 
The model can be tested at collider experiments 
 through flavor-violating processes involving sparticles. 
The flavor structure among sparticle, once observed, 
 gives us a clue to deep understanding of the origin of 
 Yukawa coupling hierarchy.

\thispagestyle{empty}

\newpage

\addtocounter{page}{-1}
\setcounter{footnote}{0}
\baselineskip 18pt
%
\section{Introduction}

The gauge hierarchy problem has been the main motivation 
 for physics beyond the Standard Model. 
One notable solution to this problem is offered by 
 Randall-Sundrum (RS) model \cite{RS}, 
 which connects 4D Planck scale and the electroweak scale 
 by means of 5D warped geometry. 
The RS setup also accommodates a natural explanation to 
 the Yukawa coupling hierarchy in its extension with matters 
 in the 5D bulk \cite{RS Yukawa}.  
This bulk matter RS model solves the Yukawa hierarchy problem 
 with the following common structure. 
We put Higgs on the infrared (IR) brane and fermions in the bulk. 
With non-hierarchical 5D Dirac masses, we localize the heavy fermions 
 towards the IR brane and the light ones towards the ultraviolet (UV) brane.
The wavefunction overlaps between the Higgs field and the fermion fields 
 give rise to the Yukawa coupling hierarchy. 
In spite of their elegant solutions to the gauge hierarchy problem 
 and the Yukawa hierarchy mystery, the RS models are under severe 
 experimental constraints concerning the Kaluza-Klein (KK) scale
 due to the flavor-changing neutral currents caused by the KK states. 
Apart from the proton decay problem, the most stringent constraint 
 comes from the data on $K^{0}-\bar{K}^{0}$ mixing 
 and the 1st KK gluon mass should be $\gtrsim 21$ TeV
 \cite{constraints on RS}. 
This constraint thus spoils their solution to the gauge hierarchy
 problem as well as their experimental accessibility 
 at the Large Hadron Collider (LHC).  
(However imposing some flavor symmetry softens 
 the KK scale bound, see \cite{flavor symmetry}.)

In this paper, we study the supersymmterization of bulk matter RS model \cite{SUSY RS}.
This model resorts to supersymmetry (SUSY) to solve the gauge hierarchy problem
 while maintaining the natural explanation of bulk matter RS model 
 to the Yukawa hierarchy; 
the KK scale can be much higher than the electroweak scale and 
 SUSY fills the gap between them. 
We especially investigate 5D Minimal SUSY Standard Model (MSSM) 
 in RS spacetime,  where we allow the gauge superfields 
 to propagate in the bulk but localize Higgs superfields and 
 the SUSY breaking sector on the IR brane, 
 and the matter superfields are laid in the bulk 
 with various 5D profiles.

All phenomenological SUSY models require a realistic SUSY breaking
 mechanism in harmony with experimental bounds.
Our model naturally incorporates gaugino-mediated SUSY breaking mechanism 
 as the matter superfields and SUSY breaking sector are (partly) 
 separated by the 5th dimension while the gauge superfields 
 couple to the SUSY breaking sector without suppression and 
 mediate its effects to matter sector. 
At the same time, it explains the Yukawa coupling hierarchy 
 through 5D localization of matter fields.
Due to this structure, the sequestering is always incomplete;
 the superfields of 3rd generation particles lean towards the IR brane 
 so that they couple to the SUSY breaking sector with less suppression.
We thus have a unique pattern of flavor-violating soft mass terms 
 that are related to the origin of the Yukawa hierarchy.
Current experimental bounds on flavor violations give strong 
 constraints on the model and allow us to make predictions on 
 the sparticle mass spectrum as well as on its flavor-violating effects.
A similar setup was proposed in \cite{RS flavorful} 
 as a 5D realization of ``flavorful supersymmetry" \cite{flavorful}.

Our model is a simple 5D extension of the MSSM in the RS spacetime, 
 but has the ability to simultaneously provide a viable SUSY breaking mechanism 
 and explain the hierarchical structure of Yukawa couplings. 
Since the sparticle mass spectrum and the Yukawa coupling hierarchy 
 are rooted on the same 5D setup, this model has a strong predictive 
 power on both of flavor-conserving and flavor-violating 
 soft SUSY breaking terms.

In the next section, we write down a general form of MSSM in the bulk 
 of RS spacetime equipped with gaugino-mediated SUSY breaking.
In section 3, we assume that all couplings in 5D theory are of $O(1)$, 
 and any hierarchical structure in 4D theory originates from 5D geometry.
Based on this assumption, we determine 5D disposition of 
 matter superfields from the observed fermion masses, 
 Cabibbo-Kobayashi-Maskawa (CKM) matrix and neutrino oscillation data. 
In section 4, we make general remarks on the SUSY breaking mass spectrum.
In section 5, we discuss the difference between our model 
 and minimal flavor violation. 
In section 6, we calculate a sample of mass spectra and study
 experimental bounds on them.
In section 7, we discuss a signature of the model, that is, 
 unusual next-to-lightest sparticle (NLSP) and its flavor-violating decay. 
The last section is devoted for conclusion.

\section{Setup}

We consider 5D warped spacetime with the metric \cite{RS}: 
\begin{equation}
{\rm d}s^{2} \ = \ e^{-2k \vert y \vert} \eta_{\mu \nu} {\rm d}x^{\mu} {\rm d}x^{\nu} - {\rm d}y^{2} \ , 
\end{equation}
where $y$ is the 5th dimension compactified on the orbifold $S^{1}/Z_{2} : -\pi R \leq y \leq \pi R $ ,
and $k$ is the AdS curvature that is of the same order as the 5D Planck scale $M_{5}$.
Assuming that the warp factor, $e^{-k R \pi}$, is much smaller than $1$,
we have the following relation for $k$ and $M_{5}$ : 
\begin{equation}
M_{*}^{2} \ = \ \frac{M_{5}^{3}}{k} ( 1-e^{-2k R \pi} ) \ \simeq \ \frac{M_{5}^{3}}{k} \ ,
\end{equation}
where $M_{*}$ is the 4D reduced Planck mass.
This relation implies $k \sim M_{5} \sim M_{*}$ .
We put a UV brane at $y=0$ and an IR brane at $y=\pi R$.
The fundamental scale on the UV brane is $M_{5}$,
while that on the IR brane is $M_{5} e^{-kR\pi}$.
Note that in our model, $M_{5} e^{-kR\pi}$ is not necessarily at TeV scale, 
but is at an intermediate scale between $M_{*}$ and TeV.

All MSSM superfields reside in the bulk.
However, for simplicity, the Higgs superfields are assumed 
 to be localized on the IR brane.
We adopt Polonyi model \cite{Polonyi} for the SUSY breaking sector 
 and introduce a gauge-singlet chiral superfield $X$ on the IR brane, 
 whose F-component develops VEV to break supersymmetry. 
We take advantage of Giudice-Masiero mechanism \cite{GM}, 
 namely, we impose an appropriate R-symmetry to 
 forbid the SUSY-conserving $\mu$-term and force the $\mu$-term to
 arise from SUSY breaking effects. 
In this paper, we assign the following R-charges 
 to $X$ and the MSSM superfields: 
\begin{eqnarray*}
X:0, \ \ \ H_{u/d}:0, \ \ \ Q_{i}/U_{i}/D_{i}/L_{i}/E_{i}:+1,
\end{eqnarray*}
where $H_{u}, \ H_{d}, \ Q_{i}, \ U_{i}, \ D_{i}, \ L_{i}, \ E_{i}$, 
 respectively, denote the chiral superfields of 
 up-type Higgs doublet, down-type Higgs doublet, SU(2) doublet quark, 
 singlet up-type quark, singlet down-type quark, 
 doublet lepton, singlet charged lepton. 
Note that the above assignment permits higher dimensional
 superpotential for light neutrino Majorana masses.

The 5D bulk action is described with 5D ${\cal N}=1$ gauge 
 multiplets and matter hyper-multiplets. 
We use 4D superfield formalism extended with the 5th dimension $y$, 
 following \cite{SUSY RS}.

An off-shell 5D ${\cal N}=1$ gauge multiplet consists 
 of a 5D gauge field $A_{M} \ (M=0,1,2,3,5)$, 
 two 4D Weyl spinors $\lambda_{1}, \lambda_{2}$, a real scalar $\Sigma$,
 a real auxiliary field $D$ and a complex auxiliary field $F$, 
 all of which transform as the adjoint representation of some gauge group. 
They are composed into one 4D ${\cal N}=1$ gauge superfield $V$ 
 and one 4D ${\cal N}=1$ chiral superfield $\chi$ that are 
\begin{eqnarray*}
V &=& -\theta \sigma^{\mu} \bar{\theta} A_{\mu} 
 - i \bar{\theta}^{2} \theta \lambda_{1} 
 + i \theta^{2} \bar{\theta} \bar{\lambda}_{1} 
 + \frac{1}{2} \bar{\theta}^{2} \theta^{2} D \ ,
\\ \chi &=& \frac{1}{\sqrt{2}} ( \Sigma + i A_{5} ) 
 + \sqrt{2} \theta \lambda_{2} + \theta^{2} F \ .
\end{eqnarray*}
Under $Z_{2}$ parity: $y \rightarrow -y$, they transform as 
\begin{eqnarray*}
V \ \rightarrow \ V \ , \ \ \ \ \ \chi \ \rightarrow \ -\chi \ .
\end{eqnarray*}
The action for 5D ${\cal N}=1$ gauge multiplets is given by
\begin{eqnarray}
S_{5D \, gauge} &=& \int {\rm d}y \int {\rm d}^{4}x \ e^{-4k \vert y \vert} \ 
\left[ \ \frac{1}{4 (g^{a}_{5})^{2}} \int {\rm d}^{2} \theta e^{k \vert y \vert} \ {\rm tr} \left\{ \ (e^{\frac{3}{2} k \vert y \vert} W^{a \, \alpha}) (e^{\frac{3}{2} k \vert y \vert} W^{a}_{\alpha})  
\ + \ {\rm h.c.} \ \right\} \right. \nonumber
\\ & & \left. \ + \ \frac{1}{(g^{a}_{5})^{2}} \int {\rm d}^{4} \theta e^{2k \vert y \vert} \ 
{\rm tr} \left\{ \ ( \sqrt{2} \partial_{y} + \chi^{a \, \dagger} ) e^{-V}  ( - \sqrt{2} \partial_{y} + \chi^{a} ) e^{V} 
\ - \ (\partial_{y} e^{-V}) (\partial_{y} e^{V}) \right\} \right] \ , \nonumber \\
\end{eqnarray}
where $a$ labels gauge groups and $W^{a \, \alpha}$ denotes the field strength of $V^{a}$ in 4D flat spacetime.
When the unitary gauge, $A^{a}_{5}=0$, is chosen,
only $V^{a}$ has a massless mode in 4D picture.
This mode has no dependence on $y$ and will be written as $V_{0}(x,\theta,\bar{\theta})$.

A 5D ${\cal N}=1$ hypermultiplet is expressed in terms of two 4D ${\cal N}=1$ chiral superfields $\Phi, \Phi^{c}$ 
that are in conjugate representations of some gauge group.
We assume that the former is $Z_{2}$-even and the latter $Z_{2}$-odd.
Taking the basis of diagonal bulk mass, we have the following action for 5D ${\cal N}=1$ hyper-multiplets:
\begin{eqnarray}
S_{5D \, chiral} &=& \int {\rm d}y \int {\rm d}^{4}x e^{-4k \vert y \vert} \ \left[ \ \int {\rm d}^{4} \theta e^{2k \vert y \vert} \ \right.
 ( \Phi_{i}^{\dagger} e^{-V} \Phi_{i} \ + \ \Phi_{i}^{c} e^{V} \Phi_{i}^{c \, \dagger} ) \nonumber
\\ & & \left. \ + \ \int {\rm d}^{2} \theta e^{k \vert y \vert} \ \Phi_{i}^{c} \{ \partial_{y} - \chi/\sqrt{2} - (3/2-c_{i}) k \} \Phi_{i} \ + \ {\rm h.c.} \ \right] \ ,
\end{eqnarray}
where $i$ is a flavor index and $c_{i}$ denotes the 5D bulk mass in unit of AdS curvature $k$.
Only $\Phi_{i}$ has a massless mode in 4D picture, which will be written as \ $\phi_{i}(x, \theta) e^{(3/2-c_{i}) k \vert y \vert}$.

Let us write down the low-energy 4D effective action of 
 the fields in the bulk, which is described with the massless modes 
 of 5D ${\cal N}=1$ gauge multiplets and 
 5D ${\cal N}=1$ matter hyper-multiplets. 
After integrating over $y$, we obtain the following 4D effective action:
\begin{eqnarray}
S_{4D \, eff.} = \int {\rm d}^{4}x \ \left[ \ \frac{2 \pi R}{4
g_{5}^{a \, 2}} \int {\rm d}^{2}\theta \ W^{a \alpha} W^{a}_{\alpha} \
+ \ {\rm h.c.} \right. 
+ \left. \int {\rm d}^{4}\theta \ 2 \frac{ e^{(1-2c_{i})kR\pi}-1 }{(1-2c_{i})k} \ \phi_{i}^{\dagger} \ e^{-V} \phi_{i} \ \right] \ , \nonumber \\ 
\end{eqnarray}
where the dimensionful 5D gauge coupling, $g_{5}^{a}$, 
 is connected to 4D gauge coupling $g_{4}^{a}$ by the relation:  
 $g_{5}^{a} = \sqrt{2\pi R} g_{4}^{a}$.

Next we consider the theory on the IR brane.
The IR scale, $M_{5} e^{-kR\pi}$, is a free parameter of the model
 and is only assumed at an intermediate scale between the 5D Planck and the electroweak scales.

On the IR brane, we introduce Polonyi model for SUSY breaking:  
\begin{eqnarray}
S_{IR} &\supset& \int {\rm d}^{4}x \ \left[ \int {\rm d}^{4}\theta \ e^{-2kR\pi} \ ( \ X^{\dagger}X \ + \ ... \ ) \ + \ \int {\rm d}^{2}\theta \ \mu_{X}^{2} X \ + \ {\rm h.c.} \right] \ .
\end{eqnarray}
 where the ``..." term is for stabilizing the scalar potential of $X$ at the origin.
$\mu_{X}$ satisfies
\begin{eqnarray}
e^{-2kR\pi} \mu_{X}^{2} &\sim& M_{5}e^{-kR\pi} \times {\rm TeV} \ ,
\end{eqnarray}
 which is equivalent to
\begin{eqnarray}
\frac{ \mu_{X} }{ M_{5} } &\sim& \sqrt{ \frac{ {\rm TeV} }{ M_{5}e^{-kR\pi} } } \ ,
\end{eqnarray}
 so that it gives rise to gaugino masses at TeV scale through the VEV of $F_{X}$.
Note that the scale of $\mu_{X}$ is between the 5D Planck and the IR scales.
This scale is put in by hand, as in tree-level SUSY breaking models,
 or is generated through a dynamical SUSY breaking mechanism \cite{DSB},
 of which the Polonyi term (6) is the effective theory.
We additionally assume that only the SUSY breaking term explicitly breaks the R-symmetry.

Other terms on the IR brane are listed below:

MSSM term:
\begin{eqnarray}
S_{IR} &\supset& \int {\rm d}^{4}x \ \left[ \int {\rm d}^{4}\theta \ e^{-2kR\pi} \ \left\{ \ H_{u}^{\dagger} e^{-V} H_{u} \ + \ H_{d}^{\dagger} e^{-V} H_{d} \ \right\} \right. \nonumber
\\ &+& \int {\rm d}^{2}\theta \ e^{-3kR\pi} \ \left\{ \ e^{(3-c_{i}-c_{j})kR\pi} \ \frac{(y_{u})_{ij}}{M_{5}} H_{u} U_{i} Q_{j} 
\ + \ e^{(3-c_{k}-c_{l})kR\pi} \ \frac{(y_{d})_{kl}}{M_{5}} H_{d} D_{k} Q_{l} \ \right\} \ + \ {\rm h.c.} \nonumber
\\ &+& \left. \int {\rm d}^{2}\theta \ e^{-3kR\pi} \ e^{(3-c_{m}-c_{n})kR\pi} \ \frac{(y_{e})_{mn}}{M_{5}} H_{d} E_{m} L_{n} \ + \ {\rm h.c.} \ \right] \ .
\end{eqnarray}

Gaugino mass term:
\begin{eqnarray}
S_{IR} &\supset& \int {\rm d}^{4}x \ \left[ \int {\rm d}^{2}\theta \ d_{a} \frac{X}{M_{5}} W^{a \, \alpha} W^{a}_{\alpha} \ + \ {\rm h.c.} \right] \ .
\end{eqnarray}

Higgs SUSY breaking term:
\begin{eqnarray}
S_{IR} &\supset& \int {\rm d}^{4}x \ \left[ \int {\rm d}^{4}\theta \ e^{-2kR\pi} \ \left\{ \ d_{mu} \frac{X^{\dagger}}{M_{5}} H_{u} H_{d} 
\ + \ d_{bmu} \frac{X^{\dagger}X}{M_{5}^{2}} H_{u} H_{d} \ + \ {\rm h.c.} \ \right\} \right. \nonumber 
\\ &+& \int {\rm d}^{4}\theta \ e^{-2kR\pi} \ \left\{ \ d_{uA} \frac{X+X^{\dagger}}{M_{5}} H_{u}^{\dagger} H_{u} \ + \
d_{u0} \frac{X^{\dagger}X}{M_{5}^{2}} H_{u}^{\dagger} H_{u} \right. \nonumber
\\ & & \left. \left. \ \ \ + \ d_{dA} \frac{X+X^{\dagger}}{M_{5}} H_{d}^{\dagger} H_{d} \ + \
d_{d0} \frac{X^{\dagger}X}{M_{5}^{2}} H_{d}^{\dagger} H_{d} \ \right\} \right] \ .
\end{eqnarray}

Matter soft mass term:
\begin{eqnarray}
S_{IR} &\supset& \int {\rm d}^{4}x \ \left[ \int {\rm d}^{4}\theta \ e^{-2kR\pi} \ e^{(3-c_{i}-c_{j})kR\pi} \ 
\left\{ \ d_{Q1 \, ij} \frac{X+X^{\dagger}}{M^{2}_{5}} \ Q_{i}^{\dagger} Q_{j} \ + \ d_{Q2 \, ij} \frac{X^{\dagger} X}{M^{3}_{5}} \ Q_{i}^{\dagger} Q_{j} \ \right\} \right] \nonumber
\\ & + & ( \ Q \ \rightarrow \ U, \ D, \ L, \ E \ ) \ .
\end{eqnarray}

A-term-generating term:
\begin{eqnarray}
S_{IR} &\supset& \int {\rm d}^{4}x \ \left[ \int {\rm d}^{2}\theta \ e^{-3kR\pi} \ \left\{ \ e^{(3-c_{i}-c_{j})kR\pi} \ \frac{(a_{u})_{ij}}{M_{5}^{2}} X H_{u} U_{i} Q_{j} 
\ + \ e^{(3-c_{k}-c_{l})kR\pi} \ \frac{(a_{d})_{kl}}{M_{5}^{2}} X H_{d} D_{k} Q_{l} \ \right. \right. \nonumber
\\ & + & \left. \left. e^{(3-c_{m}-c_{n})kR\pi} \ \frac{(a_{e})_{mn}}{M_{5}^{2}} X H_{d} E_{m} L_{n} \ \right\} \ + \ {\rm h.c.} \ \right] \ .
\end{eqnarray}
We omitted brane kinetic terms because they only affect 
 the overall normalization of the fields and are irrelevant 
 to the point of our model.

We normalize $X, \ H_{u}, \ H_{d}, \ Q_{i}, \ U_{i}, \ D_{i}, \ L_{i}, \ E_{i}$ to make their kinetic terms of the 4D effective theory canonical. 
This is done by the following rescaling:
\begin{eqnarray}
X \ \rightarrow \ \tilde{X}=e^{-kR\pi}X,
\ \ H_{u} \ \rightarrow \ \tilde{H}_{u}=e^{-kR\pi} H_{u}, 
\ \ H_{d} \ \rightarrow \ \tilde{H}_{d}=e^{-kR\pi} H_{d}, \nonumber \\
\phi_{i} \ \rightarrow \ \tilde{\phi}_{i}=\sqrt{ 2 \frac{e^{(1-2c_{i})kR\pi}-1}{(1-2c_{i})k} } \ \phi_{i},
\end{eqnarray}
where $\phi_{i}$ denotes $Q_{i}, \ U_{i}, \ D_{i}, \ L_{i}$ or $E_{i}$.
Then the MSSM term becomes
\begin{eqnarray}
S_{IR} &\supset& \int {\rm d}^{4}x \ \left[ \ \int {\rm d}^{4}\theta \ \left\{ \ \tilde{H}_{u}^{\dagger} e^{-V} \tilde{H}_{u} 
\ + \ \tilde{H}_{d}^{\dagger} e^{-V} \tilde{H}_{d} \ \right\} \right. \nonumber
\\ &+& \int {\rm d}^{2}\theta \ \left\{ \ \sqrt{ \frac{1-2c_{i}}{2\{1-e^{-(1-2c_{i})kR\pi}\}} } \sqrt{ \frac{1-2c_{j}}{2\{1-e^{-(1-2c_{j})kR\pi}\}} } 
\frac{k}{M_{5}} (y_{u})_{ij} \ \tilde{H}_{u} \tilde{U}_{i} \tilde{Q}_{j} \right. \nonumber
\\ &+& \sqrt{ \frac{1-2c_{k}}{2\{1-e^{-(1-2c_{k})kR\pi}\}} } \sqrt{ \frac{1-2c_{l}}{2\{1-e^{-(1-2c_{l})kR\pi}\}} } 
\frac{k}{M_{5}} (y_{d})_{kl} \ \tilde{H}_{d} \tilde{D}_{k} \tilde{Q}_{l} \nonumber
\\ &+& \left. \left. \sqrt{ \frac{1-2c_{m}}{2\{1-e^{-(1-2c_{m})kR\pi}\}} } \sqrt{ \frac{1-2c_{n}}{2\{1-e^{-(1-2c_{n})kR\pi}\}} }
\frac{k}{M_{5}} (y_{e})_{mn} \ \tilde{H}_{d} \tilde{E}_{m} \tilde{L}_{n} \ \right\} \ + \ {\rm h.c.} \right] .
\end{eqnarray}

The gaugino mass term will be
\begin{eqnarray}
S_{IR} &\supset& \int {\rm d}^{4}x \ \left[ \int {\rm d}^{2}\theta \ d_{a} \frac{\tilde{X}}{M_{5}e^{-kR\pi}} W^{a \, \alpha} W^{a}_{\alpha} \ + \ {\rm h.c.} \right] \ .
\end{eqnarray}

The Higgs SUSY breaking term will be
\begin{eqnarray}
S_{IR} &\supset& \int {\rm d}^{4}x \ \left[ \int {\rm d}^{4}\theta \ \left\{ \ d_{mu} \frac{\tilde{X}^{\dagger}}{M_{5}e^{-kR\pi}} \tilde{H}_{u} \tilde{H}_{d}
 \ + \ d_{bmu} \frac{\tilde{X}^{\dagger}\tilde{X}}{M_{5}^{2}e^{-2kR\pi}} \tilde{H}_{u} \tilde{H}_{d} \ + \ {\rm h.c.} \ \right. \right. \nonumber 
\\ &+& \ d_{uA} \frac{\tilde{X}+\tilde{X}^{\dagger}}{M_{5}e^{-kR\pi}} \tilde{H}_{u}^{\dagger} \tilde{H}_{u} \ + \
d_{u0} \frac{\tilde{X}^{\dagger}\tilde{X}}{M_{5}^{2}e^{-2kR\pi}} \tilde{H}_{u}^{\dagger} \tilde{H}_{u} \nonumber
\\ & & \left. \left. \ \ \ + \ d_{dA} \frac{\tilde{X}+\tilde{X}^{\dagger}}{M_{5}e^{-kR\pi}} \tilde{H}_{d}^{\dagger} \tilde{H}_{d} \ + \
d_{d0} \frac{\tilde{X}^{\dagger}\tilde{X}}{M_{5}^{2}e^{-2kR\pi}} \tilde{H}_{d}^{\dagger} \tilde{H}_{d} \ \right\} \right] \ .
\end{eqnarray}

The matter soft mass term will be
\begin{eqnarray}
S_{IR} &\supset& \int {\rm d}^{4}x \ \left[ \int {\rm d}^{4}\theta \ \sqrt{ \frac{1-2c_{i}}{2\{1-e^{-(1-2c_{i})kR\pi}\}} } \sqrt{ \frac{1-2c_{j}}{2\{1-e^{-(1-2c_{j})kR\pi}\}} } \ \frac{k}{M_{5}} \ \times \right. \nonumber
\\
& & \ \ \left. \left\{ \ d_{Q1 \, ij} \frac{\tilde{X}+\tilde{X}^{\dagger}}{M_{5}e^{-kR\pi}} \ \tilde{Q}_{i}^{\dagger} \tilde{Q}_{j} \ + \ 
d_{Q2 \, ij} \frac{\tilde{X}^{\dagger} \tilde{X}}{M^{2}_{5} e^{-2kR\pi}} \ \tilde{Q}_{i}^{\dagger} \tilde{Q}_{j} \  \right\} \right] \nonumber
\\ &+& ( \ \tilde{Q} \ \rightarrow \ \tilde{U}, \ \tilde{D}, \ \tilde{L}, \ \tilde{E} \ ) \ .
\end{eqnarray}

The A-term-generating term will be
\begin{eqnarray}
S_{IR} &\supset& \int {\rm d}^{4}x \ \left[ \int {\rm d}^{2}\theta \ \left\{ \ \sqrt{ \frac{1-2c_{i}}{2\{1-e^{-(1-2c_{i})kR\pi}\}} } \sqrt{ \frac{1-2c_{j}}{2\{1-e^{-(1-2c_{j})kR\pi}\}} } 
\frac{k}{M_{5}} \frac{(a_{u})_{ij}}{M_{5} e^{-kR\pi}} \ \tilde{X} \tilde{H}_{u} \tilde{U}_{i} \tilde{Q}_{j} \right. \right. \nonumber
\\ &+& \sqrt{ \frac{1-2c_{k}}{2\{1-e^{-(1-2c_{k})kR\pi}\}} } \sqrt{ \frac{1-2c_{l}}{2\{1-e^{-(1-2c_{l})kR\pi}\}} } 
\frac{k}{M_{5}} \frac{(a_{d})_{kl}}{M_{5} e^{-kR\pi}} \ \tilde{X} \tilde{H}_{d} \tilde{D}_{k} \tilde{Q}_{l} \nonumber
\\ &+& \left. \left. \sqrt{ \frac{1-2c_{m}}{2\{1-e^{-(1-2c_{m})kR\pi}\}} } \sqrt{ \frac{1-2c_{n}}{2\{1-e^{-(1-2c_{n})kR\pi}\}} }
\frac{k}{M_{5}} \frac{(a_{e})_{mn}}{M_{5} e^{-kR\pi}} \ \tilde{X} \tilde{H}_{d} \tilde{E}_{m} \tilde{L}_{n} \ \right\} \ + \ {\rm h.c.} \ \right] \ . \nonumber \\
\end{eqnarray}

We introduce light neutrino masses by simply writing down higher dimensional operators on the IR brane, namely,
\begin{eqnarray}
S_{IR} &\supset& \int {\rm d}^{4}x \int {\rm d}^{2}\theta \ e^{-3kR\pi} \ e^{(3-c_{p}-c_{q})kR\pi} \ 
(Y_{\nu})_{pq} \frac{L_{p} H_{u} L_{q} H_{u}}{M_{5}} \ + \ {\rm h.c.} \ \nonumber
\\ &=& \int {\rm d}^{4}x \int {\rm d}^{2}\theta \ \sqrt{ \frac{1-2c_{p}}{2\{1-e^{-(1-2c_{p})kR\pi}\}} } \sqrt{ \frac{1-2c_{q}}{2\{1-e^{-(1-2c_{q})kR\pi}\}} } \ 
(Y_{\nu})_{pq} \ \frac{\tilde{L}_{p} \tilde{H}_{u} \tilde{L}_{q} \tilde{H}_{u}}{M_{5} e^{-kR\pi}} \ + \ {\rm h.c.} \nonumber \\
\end{eqnarray}
Note that the Kaluza-Klein (KK) scale, $M_{5} e^{-kR\pi}$, 
 is related to the scale of light neutrino masses. 
Another possibility is to introduce singlet neutrino superfields 
 and adopt the seesaw mechanism \cite{seesaw}. 
In this case, the KK scale can be a free parameter of the model.

Now the MSSM Yukawa couplings are expressed as 
\begin{eqnarray}
(Y_{u})_{ij} &=& \sqrt{ \frac{1-2c_{i}}{2\{1-e^{-(1-2c_{i})kR\pi}\}} } \sqrt{ \frac{1-2c_{j}}{2\{1-e^{-(1-2c_{j})kR\pi}\}} }
\frac{k}{M_{5}} (y_{u})_{ij} \ , \nonumber
\\ 
(Y_{d})_{kl} &=& \sqrt{ \frac{1-2c_{k}}{2\{1-e^{-(1-2c_{k})kR\pi}\}} } \sqrt{ \frac{1-2c_{l}}{2\{1-e^{-(1-2c_{l})kR\pi}\}} }
\frac{k}{M_{5}} (y_{d})_{kl} \ , \nonumber
\\
(Y_{e})_{mn} &=& \sqrt{ \frac{1-2c_{m}}{2\{1-e^{-(1-2c_{m})kR\pi}\}} } \sqrt{ \frac{1-2c_{n}}{2\{1-e^{-(1-2c_{n})kR\pi}\}} }
\frac{k}{M_{5}} (y_{e})_{mn} \ ,
\end{eqnarray}
and the neutrino mass matrix $m_{\nu}$ is given by 
\begin{eqnarray}
(m_{\nu})_{pq} &=& \sqrt{ \frac{1-2c_{p}}{2\{1-e^{-(1-2c_{p})kR\pi}\}} } \sqrt{ \frac{1-2c_{q}}{2\{1-e^{-(1-2c_{q})kR\pi}\}} }
\ (Y_{\nu})_{pq} \ \frac{v_{u}^{2}}{M_{5}e^{-kR\pi}} \ .
\end{eqnarray}
The geometrical factor $\sqrt{ (1-2c) \ / \ (2\{1-e^{-(1-2c)kR\pi}\}) }$ has a unique property.
For $c < 1/2$, it is approximated by $\sqrt{ 1/2 - c }$ and is $O(1)$.
For $c > 1/2$, it is approximated by $\sqrt{ c - 1/2 } \ e^{-(c-1/2)kR\pi}$ and is exponentially suppressed.
Therefore this factor can generate the large hierarchy of the Yukawa couplings without hierarchy.
In the following, we assume that the components of 5D coupling matrices, $y_{u}, \ y_{d}, \ y_{e}, \ Y_{\nu}$, are all $O(1)$
and that the hierarchical structure of MSSM Yukawa couplings and the neutrino mass matrix arises from the following terms:
\begin{eqnarray*}
\sqrt{ \frac{1-2c_{i}}{2\{1-e^{-(1-2c_{i})kR\pi}\}} } \sqrt{ \frac{1-2c_{j}}{2\{1-e^{-(1-2c_{j})kR\pi}\}} } \ .
\end{eqnarray*}

We define geometrical factors $\alpha_{i}$ as
\begin{equation}
\alpha_{i} \ \equiv \ \sqrt{ \frac{1-2c_{q \, i}}{2\{1-e^{-(1-2c_{q \, i})kR\pi}\}} } \ \ \ \ \ {\rm with} \ \ i=1,2,3
\end{equation}
for the $i$-th generation of SU(2) doublet quark superfields.
Similarly, we define $\beta_{i}, \ \gamma_{i}, \ \delta_{i}, \ \epsilon_{i}$
 for SU(2) singlet up-type quark, singlet down-type quark, 
 doublet lepton, singlet neutrino and singlet charged lepton, respectively.
Thus, the up-type quark Yukawa matrix $Y_{u}$, 
 the down-type Yukawa matrix $Y_{d}$ 
 and the charged lepton Yukawa matrix $Y_{e}$ 
 (in the basis of diagonal 5D bulk mass) are given by
\begin{equation}
(Y_{u})_{ij} \ \sim \ \beta_{i}\alpha_{j} \ , \ \ \ (Y_{d})_{ij} \ \sim \ \gamma_{i}\alpha_{j} \ , \ \ \ (Y_{e})_{ij} \ \sim \ \epsilon_{i}\delta_{j} \ ,
\end{equation}
and the neutrino mass matrix $m_{\nu}$ is given by
\begin{equation}
(m_{\nu})_{ij} \ \sim \ \delta_{i} \delta_{j} \ \frac{v_{u}^{2}}{M_{5}e^{-kR\pi}},   
\end{equation}
with VEV of the up-type Higgs doublet $v_u$.

\section{Yukawa coupling hierarchy from geometry}

In this section, we determine the order of the geometrical factors, 
 $\alpha_{i}, \beta_{i}, \gamma_{i}, \delta_{i}, \epsilon_{i}$,
 from the experimental data on SM fermion masses, CKM matrix 
 and the neutrino oscillation parameters.
Note that the geometrical factors must be evaluated 
 at the KK scale, $k e^{-kR\pi}$, where the 5D theory 
 is connected to the 4D effective theory. 
However, as is seen from \cite{scale dependence of}, 
 the renormalization group (RG) running changes 
 the Yukawa couplings by at most a factor $2$ and 
 CKM matrix components by at most $1.2$ 
 through the RG running from $\sim 10^{15}$ GeV to electroweak scale.
Also the neutrino mass matrix is affected only by $O(1)$ 
 through the RG running \cite{RGE for neutrinos}. 
Therefore we can estimate the order 
 of $\alpha_{i}, \beta_{i}, \gamma_{i}, \delta_{i}, \epsilon_{i}$ 
 directly from the experimental data at low energies.

We first show the model's predictions on Yukawa eigenvalues and CKM matrix.
Let us diagonalize the Yukawa matrices:
\begin{eqnarray*}
V_{u} Y_{u} U_{u}^{\dagger} &=& {\rm diag} \ ,
\\ V_{d} Y_{d} U_{d}^{\dagger} &=& {\rm diag} \ ,
\\ V_{e} Y_{e} U_{e}^{\dagger} &=& {\rm diag} \ .
\end{eqnarray*}
For successful diagonalization of the hierarchical Yukawa matrices, 
the unitary matrices, \\ $U_{u}, \ U_{d}, \ V_{u}, \ V_{d}, \ U_{e}, \ V_{e}, $ need to have the following structure: 
\begin{eqnarray}
U_{u} \ \sim \ U_{d} \ \sim \ \left(
\begin{array}{ccc}
1 & 0 & 0 \\
\alpha_{1}/\alpha_{2} & 1 & 0 \\
\alpha_{1}/\alpha_{3} & \alpha_{2}/\alpha_{3} & 1
\end{array}
\right) \ , \ \ \ 
V_{u} \ \sim \ ({\rm \alpha \rightarrow \beta}) \ , \ \ \ V_{d} \ \sim \ ({\rm \alpha \rightarrow \gamma}) \ , \nonumber
\\ U_{e} \ \sim \ ({\rm \alpha \rightarrow \delta}) \ , \ \ \ V_{e} \ \sim \ ({\rm \alpha \rightarrow \epsilon})
\end{eqnarray}
which leads to
\begin{eqnarray}
V_{u} Y_{u} U_{u}^{\dagger} &\sim& {\rm diag} \ ( \ \beta_{1}\alpha_{1}, \ \beta_{2}\alpha_{2}, \ \beta_{3}\alpha_{3} \ ) \ , \nonumber
\\ V_{d} Y_{d} U_{d}^{\dagger} &\sim& {\rm diag} \ ( \ \gamma_{1}\alpha_{1}, \ \gamma_{2}\alpha_{2}, \ \gamma_{3}\alpha_{3} \ ) \ , \nonumber
\\ V_{e} Y_{e} U_{e}^{\dagger} &\sim& {\rm diag} \ ( \ \epsilon_{1}\delta_{1}, \ \epsilon_{2}\delta_{2}, \ \epsilon_{3}\delta_{3} \ ) \ .
\end{eqnarray}
The hierarchical structure of CKM matrix $U_{CKM}$ is given by 
\begin{eqnarray}
U_{CKM} &=& U_{u} U_{d}^{\dagger} \ \sim \ \left(
\begin{array}{ccc}
1 & \alpha_{1}/\alpha_{2} & \alpha_{1}/\alpha_{3} \\
\alpha_{1}/\alpha_{2} & 1 & \alpha_{2}/\alpha_{3} \\
\alpha_{1}/\alpha_{3} & \alpha_{2}/\alpha_{3} & 1
\end{array}
\right) \ .
\end{eqnarray}

The absolute values of the CKM matrix components, 
 $\vert U_{CKM} \vert$, at electroweak scale has been measured as \cite{pdg}
\begin{eqnarray*}
& & \vert U_{CKM} [ M_{W} ] \vert \\ &=& \left(
\begin{array}{ccc}
0.97419 \pm 0.00022 & 0.2257 \pm 0.0010 & 0.00359 \pm 0.00016 \\
0.2256 \pm 0.0010 & 0.97334 \pm 0.00023 & 0.0415 + 0.0010 -0.0011 \\
0.00874 + 0.00026 - 0.00037 & 0.0407 \pm 0.0010 & 0.999133 + 0.000044 - 0.000043
\end{array}
\right) \ .
\end{eqnarray*}
We approximate this matrix by the following formula:
\begin{equation}
\vert U_{CKM} \vert \ \simeq \ \left(
\begin{array}{ccc}
1 & \lambda & \lambda^{3} \\
\lambda & 1 & \lambda^{2} \\
\lambda^{3} & \lambda^{2} & 1
\end{array}
\right) \ \ {\rm with} \ \ \lambda = 0.22 \ .
\end{equation}
To discuss the neutrino mass matrix, 
 we adopt the tri-bi-maximal mixing matrix \cite{U_TBM}
 (which gives almost the best fit in the neutrino oscillation data):
\begin{eqnarray*}
U_{MNS} &=& \left(
\begin{array}{ccc}
\sqrt{\frac{2}{3}} & \sqrt{\frac{1}{3}} & 0 \\
-\sqrt{\frac{1}{6}} & \sqrt{\frac{1}{3}} &  \sqrt{\frac{1}{2}} \\
-\sqrt{\frac{1}{6}} & \sqrt{\frac{1}{3}} & -\sqrt{\frac{1}{2}}
\end{array}
\right)
\end{eqnarray*}
and the following data on neutrino mass squared differences \cite{pdg}:
\begin{eqnarray*}
\Delta m_{21}^{2} &=& 7.59 \pm 0.20 \times 10^{-5} \ {\rm eV}^{2}, \ \ \ \ \ 
\vert \Delta m_{32}^{2} \vert \ = \  2.43 \pm 0.13 \times 10^{-3} \ {\rm eV}^{2}.
\end{eqnarray*}
Also we assume that the mass of the lightest neutrino is negligible,
 for simplicity.
Then the neutrino mass matrix, 
 $ U_{MNS} \ {\rm diag}(m_{\nu 1}, m_{\nu 2}, m_{\nu 3}) U_{MNS}^{\dagger} $ ,
is given by
\begin{eqnarray}
U_{MNS} \ {\rm diag}( m_{\nu 1},  m_{\nu 2},  m_{\nu 3}) 
 U_{MNS}^{\dagger} &=& \left(
\begin{array}{ccc}
0.29 & 0.29 & 0.29 \\
0.29 & 2.8 & -2.2 \\
0.29 & -2.2 & 2.8
\end{array}
\right) \times 10^{-11} \ {\rm GeV} 
\end{eqnarray}
for the normal hierarchy case, while 
\begin{eqnarray}
U_{MNS} \ {\rm diag}( m_{\nu 1}, m_{\nu 2}, m_{\nu 3} ) 
 U_{MNS}^{\dagger} &=& \left(
\begin{array}{ccc}
4.9 & 0.026 & 0.026 \\
0.026 & 2.5 & 2.5 \\
0.026 & 2.5 & 2.5
\end{array}
\right) \times 10^{-11} \ {\rm GeV} 
\end{eqnarray}
for the inverted hierarchy case.

Now we are ready to compare the model parameters with the experimental data
 and estimate the order of 
 $\alpha_{i}, \ \beta_{i}, \ \gamma_{i}, \ \delta_{i}, \ \epsilon_{i}$.
For Yukawa eigenvalues, we simply have 
\begin{eqnarray}
\beta_{1}\alpha_{1} &\sim& m_{u}/v \sin \beta \ , \ \ \ \beta_{2}\alpha_{2} \ \sim \ m_{c}/v \sin \beta \ , \ \ \ \beta_{3}\alpha_{3} \ \sim \ m_{t}/v \sin \beta \ ,
\\ \gamma_{1}\alpha_{1} &\sim& m_{d}/v \cos \beta \ , \ \ \ \gamma_{2}\alpha_{2} \ \sim \ m_{s}/v \cos \beta \ , \ \ \ \gamma_{3}\alpha_{3} \ \sim \ m_{b}/v \cos \beta \ ,
\\ \epsilon_{1}\delta_{1} &\sim& m_{e}/v \cos \beta \ , \ \ \ \epsilon_{2}\delta_{2} \ \sim \ m_{\mu}/v \cos \beta \ , \ \ \ \epsilon_{3}\delta_{3} \ \sim \ m_{\tau}/v \cos \beta \ .
\end{eqnarray}
Since the top Yukawa coupling is of $\sim 1$, 
 we have $\alpha_{3}\beta_{3} \sim 1$, which leads to 
\begin{equation}
\alpha_{3} \ \sim \ \beta_{3} \ \sim \ 1 \ .
\end{equation}
Comparing (28) with (29), we find 
\begin{equation}
\alpha_{1} \ \sim \ \lambda^{3} \ , \ \ \alpha_{2} \ \sim \ \lambda^{2}.
\end{equation}
We then have 
\begin{eqnarray}
\beta_{1} & \sim & \lambda^{-3} \ m_{u}/v \sin \beta \ , \ \ \beta_{2} \ \sim \ \lambda^{-2} \ m_{c}/v \sin \beta\ ,
\\ \gamma_{1} & \sim & \lambda^{-3} \ m_{d}/v \cos \beta \ , \ \ \gamma_{2} \ \sim \ \lambda^{-2} \ m_{s}/v \cos \beta \ ,
\ \ \gamma_{3} \ \sim \ m_{b}/v \cos \beta \ .
\end{eqnarray}

Next compare the matrix (25) with the observed neutrino mass matrix.
For the normal hierarchy case, it is possible to reproduce 
 the hierarchical structure of the neutrino mass matrix by adjusting 
\begin{eqnarray}
3 \delta_{1} &\sim& \delta_{2} \ \sim \ \delta_{3}  
\end{eqnarray}
with the factor $3$ coupling of the 5D theory. 
On the other hand, for the inverted hierarchy case, 
 we cannot reproduce the neutrino mass matrix 
 with ${\cal O}(1)$ couplings. 
The situation gets worse if we consider non-negligible mass of the lightest neutrino.
Therefore, the model favors the normal hierarchy of neutrino masses 
 with the relation (39). 
We estimate $\epsilon_{i}$ from the relation (39) as  
\begin{equation}
\epsilon_{1} \ \sim \ 3 \ \delta_{3}^{-1} \ m_{e}/v \cos \beta \ , \ \ \epsilon_{2} \ \sim \ \delta_{3}^{-1} \ m_{\mu}/v \cos \beta \ , \ \ 
\epsilon_{3} \ \sim \ \delta_{3}^{-1} \ m_{\tau}/v \cos \beta \ .
\end{equation}

Finally, we refer to the connection between the light neutrino mass scale and the KK scale.
If the neutrino mass arises from higher dimensional superpotential, as in (25),
the two scales are related through the following formula:
\begin{eqnarray}
\delta_{3}^{2} \ \frac{v_{u}^{2}}{M_{5} e^{-kR\pi}} &\sim& 3 \times 10^{-11} \ {\rm GeV} \ .
\end{eqnarray}
Based on the relation above, we can estimate the KK scale 
 from the value of $\delta_{3}$.

\section{Two origins of soft SUSY breaking terms}

In this model, SUSY breaking terms have two origins. 
One is contact terms between the SUSY breaking sector 
 and the MSSM sector on the IR brane 
 (gravity mediation contributions) \cite{gravity mediation}.
The other is radiative corrections, in particular, 
 the renormalization group effects from gaugino soft masses 
 (gaugino mediation contributions) \cite{gaugino mediation}.
For the superpartners of matter particles,
 the former induce flavor-violating soft terms 
 while the latter mainly generate flavor-diagonal terms
Due to the model's structure, the gravity mediation contributions 
 are related to the 5D disposition of matter superfields 
 that gives rise to the Yukawa coupling hierarchy.

As is argued in \cite{Chacko}, 
when the square root of space-like momentum,
$p \equiv \sqrt{-p^{2}} $,
is larger than the KK scale, $ke^{-kR\pi}$, 
5D gaugino propagator connecting the UV and 
 the IR branes is suppressed by the factor 
\begin{eqnarray*}
\exp[ \ - p / (ke^{-kR\pi}) \ ] \ .
\end{eqnarray*}
Since the integral of loop momentum is done with Euclidized momentum, 
$l_{E}^{2} = -l^{2}$,
loop diagrams containing gaugino propagators between the two branes
are also exponentially suppressed
when the range of integral is limited to $[O(k e^{-kR\pi}), \ \infty)$, 
that is, when the renormalization scale is around the KK scale.
Matter superfields confined on the UV brane receive SUSY breaking effects through loop diagrams 
involving gaugino propagators in the bulk and gaugino mass on the IR.
Hence we argue that,
at the scale of $ke^{-kR\pi}$,
matter SUSY particles \textit{in the bulk} gain SUSY breaking mass only through the contact terms on the IR brane,
and radiative corrections through gauginos are negligible.
When $p < ke^{-kR\pi}$, 5D gaugino propagator approaches 
 to the 4D one divided by $\pi R$, and matter SUSY particles 
 receive SUSY breaking effects through gaugino radiative corrections
 just as in 4D MSSM.
Based on the discussions above, we calculate the SUSY breaking mass spectrum 
 in the following way:
At the renormalization scale $\mu_{r} = ke^{-kR\pi}$,
 SUSY breaking terms arise only from contact terms on the IR brane (gravity mediation).
In particular, 1st generation matter sparticles that are localized 
 towards the UV brane have almost zero soft mass.
Below the scale of $ke^{-kR\pi}$, 
 the RG equations of 4D MSSM controls the mass spectrum (gaugino mediation).
Therefore we can calculate the sparticle mass spectrum 
 at the electroweak scale by solving the MSSM RG equations
 with the initial condition that, at $\mu_{r}=ke^{-kR\pi}$, SUSY breaking terms 
 be given by the IR brane contact terms.
In the rest of the paper, we denote the scale $ke^{-kR\pi}$ as $M_{cut}$.

At $\mu_{r} = M_{cut}$, the SUSY breaking terms are given as follows: 
\begin{eqnarray}
{\rm gaugino \ masses} \ \ \ M^{a}_{1/2} &=& - \ d_{a} \ 4(g_{4}^{a})^{2} \ \frac{ < F_{\tilde{X}} > }{ M_{5}e^{-kR\pi} }
\\ {\rm Higgs \ B\mu \ term} \ \ \ B\mu &=& d_{bmu} \ \frac{ \vert < F_{\tilde{X}} > \vert^{2} }{ M_{5}^{2}e^{-2kR\pi} }
\\ {\rm Higgs \ soft \ masses} \ \ \ m_{H_{u}}^{2} &=& (-d_{u0}+d_{uA}^{2}) \ \frac{ \vert < F_{\tilde{X}} > \vert^{2} }{ M_{5}^{2}e^{-2kR\pi} }
\\ m_{H_{d}}^{2} &=& (-d_{d0}+d_{dA}^{2}) \ \frac{ \vert < F_{\tilde{X}} > \vert^{2} }{ M_{5}^{2}e^{-2kR\pi} }
\\ {\rm matter \ soft \ masses} \ \ \ (m_{Q}^{2})_{ij} &=& (-d_{Q2ij}+d_{Q1ij}^{2}) \ \alpha_{i} \alpha_{j} \frac{ \vert < F_{\tilde{X}} > \vert^{2} }{ M_{5}^{2}e^{-2kR\pi} } \nonumber
\\ {\rm (Q, \ \alpha)} \ & \rightarrow & \ {\rm (U, \beta), \ (D, \gamma), \ (L, \delta), \ (E, \epsilon)}
\\ {\rm \ A-terms } \ \ \ (A_{u})_{ij} &=& -d_{uA} \ (y_{u})_{ij} \ \beta_{i} \alpha_{j} \ \frac{k}{M_{5}} \ \frac{ < F_{\tilde{X}} > }{ M_{5}e^{-kR\pi} }
\ + \ (a_{u})_{ij} \ \beta_{i} \alpha_{j} \ \frac{k}{M_{5}} \ \frac{ < F_{\tilde{X}} > }{ M_{5}e^{-kR\pi} } \nonumber
\\ &=& -d_{uA} \ (Y_{u})_{ij} \ \frac{ < F_{\tilde{X}} > }{ M_{5}e^{-kR\pi} } 
\ + \ (a_{u})_{ij} \ \beta_{i} \alpha_{j} \ \frac{k}{M_{5}} \ \frac{ < F_{\tilde{X}} > }{ M_{5}e^{-kR\pi} }
\\ (A_{d})_{ij} &=& -d_{dA} \ (Y_{d})_{ij} \ \frac{ < F_{\tilde{X}} > }{ M_{5}e^{-kR\pi} } 
\ + \ (a_{d})_{ij} \ \gamma_{i} \alpha_{j} \ \frac{k}{M_{5}} \ \frac{ < F_{\tilde{X}} > }{ M_{5}e^{-kR\pi} }
\\ (A_{e})_{ij} &=& -d_{dA} \ (Y_{e})_{ij} \ \frac{ < F_{\tilde{X}} > }{ M_{5}e^{-kR\pi} }
\ + \ (a_{e})_{ij} \ \epsilon_{i} \delta_{j} \ \frac{k}{M_{5}} \ \frac{ < F_{\tilde{X}} > }{ M_{5}e^{-kR\pi} }
\end{eqnarray}
where $\alpha_{i}, \beta_{i}, \gamma_{i}, \delta_{i}, 
 \epsilon_{i}$ are defined as in (23).
In addition, the $\mu$-term arises from the SUSY breaking effects 
 (Giudice-Masiero mechanism): 
\begin{eqnarray}
\mu &=& d_{mu} \ \frac{ < F_{\tilde{X}} > }{ M_{5}e^{-kR\pi} } \ .
\end{eqnarray}
Note that the flavor structure of matter soft masses and A-terms 
corresponds to the Yukawa coupling and neutrino mass matrix hierarchy in a unique way, 
governed by $\alpha_{i}, \ \beta_{i}, \ \gamma_{i}, \ \delta_{i}, \ \epsilon_{i}$.

We solve the MSSM RG equations from $M_{cut}$ toward low energies 
 with the initial conditions (42-50), and evaluate 
 the sparticle mass spectrum at the electroweak scale.

Finally we remark on the nature of the lightest SUSY particle (LSP) 
 and the next-to-lightest SUSY particle (NLSP) in this model. 
The gravitino mass is given by
\begin{equation}
m_{3/2} \ \simeq \ \frac{\vert <F_{\tilde{X}}> \vert}{\sqrt{3} M_{*}} \ = \ 
\frac{\vert <F_{\tilde{X}}> \vert}{\sqrt{3} M_{5}e^{-kR\pi}} \frac{M_{5}e^{-kR\pi}}{M_{*}} \ \sim \ {\rm TeV} \times e^{-kR\pi} \ ,
\end{equation}
 and thus gravitino is always LSP, as in \cite{IOY}. 
NLSP mainly consists of singlet sleptons
 whose flavor composition depends on the amount 
 of gravity mediation contributions. 
Normally, the singlet stau is lighter than smuon and
 selectron due to its large Yukawa coupling,
 but in this model, it gains large soft mass through gravity 
 mediation and may not be the lightest. 
As with other gravitino LSP scenarios, 
 NLSP is long-lived because its coupling to gravitino is suppressed by
 $1/\vert < F_{\tilde{X}} > \vert$.

\section{Comparison with Minimal Flavor Violation}

The minimal flavor violation (MFV) is the setup that  
 only SM Yukawa couplings violate flavor symmetry.
In MFV, flavor-violating soft terms are generated 
 via the MSSM RG equations involving Yukawa couplings. 
We here estimate the orders of the flavor-violating soft terms 
 generated through RG running in MFV, and compare them 
 with those via the gravity mediation in our model.
We will see that the latter
 show different patterns from the former.

We first introduce a flavor basis where $Y_{u}$ or $Y_{d}$ and $Y_{e}$ 
 are diagonalized by the following unitary matrices $U_{*}$:
\begin{eqnarray*}
U_{U} Y_{u} U_{Qu} &=& (diag.) \ , \\
U_{D} Y_{d} U_{Qd} &=& (diag.) \ , \\
U_{E} Y_{e} U_{L} &=& (diag.) \ .
\end{eqnarray*}
Note that $U_{*}$'s depend on the renormalization scale 
 as Yukawa matrices receive RG corrections.
We will estimate the orders of the changes of $U_{*}$'s through RG running.
Below is the RG equations for $Y_{u}$:
\begin{eqnarray}
\mu \frac{{\rm d}}{{\rm d} \mu} (U_{U} Y_{u} U_{Qu}) &=&
(\mu \frac{{\rm d}}{{\rm d} \mu} U_{U}) U_{U}^{\dagger} (U_{U} Y_{u} U_{Qu}) \ + \ 
U_{U} (\mu \frac{{\rm d}}{{\rm d} \mu} Y_{u}) U_{Qu} \ + \
(U_{U} Y_{u} U_{Qu}) U_{Qu}^{\dagger} (\mu \frac{{\rm d}}{{\rm d} \mu} U_{Qu}) \nonumber \\
&=& (\mu \frac{{\rm d}}{{\rm d} \mu} U_{U}) U_{U}^{\dagger} (U_{U} Y_{u} U_{Qu}) \nonumber \\
&+& \frac{1}{16 \pi^{2}} \ U_{U} \ \{ \ Y_{u} Y_{d}^{\dagger} Y_{d} + 3 Y_{u} Y_{u}^{\dagger} Y_{u} + 
3 {\rm tr}[Y_{u}^{\dagger}Y_{u}] Y_{u} + {\rm tr}[Y_{D}^{\dagger}Y_{D}] Y_{u} \nonumber \\
& & - \ (\frac{13}{15} g_{1}^{2} + 3 g_{2}^{2} + \frac{16}{3} g_{3}^{2})Y_{u} \ \} \ U_{Qu} \nonumber \\
&+& (U_{U} Y_{u} U_{Qu}) U_{Qu}^{\dagger} (\mu \frac{{\rm d}}{{\rm d} \mu} U_{Qu}) \ ,
\end{eqnarray}
where $Y_{D}$ is neutrino Dirac Yukawa coupling which appears 
 when we introduce the singlet neutrinos lighter than $M_{cut}$.
We hereafter adopt the GUT normalization for $g_{1}$. 
From (52), we see that $U_{U}Y_{u}U_{Qu}$ remains diagonal 
 during RG running when the unitary matrices 
 satisfy the following conditions,  
\begin{eqnarray}
\mu \frac{{\rm d}}{{\rm d} \mu} U_{U} &=& 0 \ , \\
\mu \frac{{\rm d}}{{\rm d} \mu} U_{Qu} &=& -\frac{1}{16 \pi^{2}} \ 
({\rm off-diagonal \ components \ of} \ Y_{d}^{\dagger} Y_{d}) \ U_{Qu} \ .
\end{eqnarray}
In the same manner, we obtain the following conditions 
 for keeping $U_{D} Y_{d} U_{Qd}$ and $U_{E} Y_{e} U_{L}$ diagonal: 
\begin{eqnarray}
\mu \frac{{\rm d}}{{\rm d} \mu} U_{D} &=& 0 \ , \\
\mu \frac{{\rm d}}{{\rm d} \mu} U_{Qd} &=& -\frac{1}{16 \pi^{2}} \ 
({\rm off-diagonal \ components \ of} \ Y_{u}^{\dagger} Y_{u}) \ U_{Qd} \ , \\
\mu \frac{{\rm d}}{{\rm d} \mu} U_{E} &=& 0 \ , \\
\mu \frac{{\rm d}}{{\rm d} \mu} U_{L} &=& -\frac{1}{16 \pi^{2}} \ 
({\rm off-diagonal \ components \ of} \ Y_{D}^{\dagger} Y_{D}) \ U_{L} \ .
\end{eqnarray}

Now that we know how $Y_{u}$-diagonal basis, 
 $Y_{d}$-diagonal basis and $Y_{e}$-diagonal basis 
 change through RG running,
 we estimate the orders of MFV effects on A-terms in these bases.
The MSSM RG equations for A-terms are given by:
\begin{eqnarray}
16 \pi^{2} \mu \frac{{\rm d}}{{\rm d}\mu} A_{u} &=& 
3 A_{u} Y_{u}^{\dagger} Y_{u} + 3 Y_{u} Y_{u}^{\dagger} A_{u} \nonumber \\
&+& A_{u} Y_{d}^{\dagger} Y_{d} + 2 Y_{u} Y_{d}^{\dagger} A_{d} \nonumber \\
&+& 2 ( \ 3 {\rm tr}[ Y_{u}^{\dagger} A_{u} ] 
- \frac{13}{15} g_{1}^{2} M^{a=1}_{1/2} - 3 g_{2}^{2} M^{a=2}_{1/2} - \frac{16}{3} g_{3}^{2} M^{a=3}_{1/2} \ ) Y_{u} \nonumber \\
&+& ( \ 3 {\rm tr}[ Y_{u}^{\dagger} Y_{u} ] - \frac{13}{15} g_{1}^{2} - 3 g_{2}^{2} - \frac{16}{3} g_{3}^{2} \ ) A_{u} \nonumber \\
&+& {\rm tr}[ Y_{D}^{\dagger} Y_{D} ] A_{u} + {\rm tr}[ Y_{D}^{\dagger} A_{D} ] Y_{u} \ ,
\\
16 \pi^{2} \mu \frac{{\rm d}}{{\rm d}\mu} A_{d} &=& 
3 A_{d} Y_{d}^{\dagger} Y_{d} + 3 Y_{d} Y_{d}^{\dagger} A_{d} \nonumber \\
&+& A_{d} Y_{u}^{\dagger} Y_{u} + 2 Y_{d} Y_{u}^{\dagger} A_{u} \nonumber \\
&+& 2 ( \ 3 {\rm tr}[ Y_{d}^{\dagger} A_{d} ] + {\rm tr}[ Y_{e}^{\dagger} A_{e} ]
- \frac{7}{15} g_{1}^{2} M^{a=1}_{1/2} - 3 g_{2}^{2} M^{a=2}_{1/2} - \frac{16}{3} g_{3}^{2} M^{a=3}_{1/2} \ ) Y_{d} \nonumber \\
&+& ( \ 3 {\rm tr}[ Y_{d}^{\dagger} Y_{d} ] + {\rm tr}[ Y_{e}^{\dagger} Y_{e} ] 
- \frac{7}{15} g_{1}^{2} - 3 g_{2}^{2} - \frac{16}{3} g_{3}^{2} \ ) A_{d} \ ,
\\
16 \pi^{2} \mu \frac{{\rm d}}{{\rm d}\mu} A_{e} &=& 
3 A_{e} Y_{e}^{\dagger} Y_{e} + 3 Y_{e} Y_{e}^{\dagger} A_{e} \nonumber \\
&+& 2 ( \ 3 {\rm tr}[ Y_{d}^{\dagger} A_{d} ] + {\rm tr}[ Y_{e}^{\dagger} A_{e} ]
- \frac{9}{5} g_{1}^{2} M^{a=1}_{1/2} - 3 g_{2}^{2} M^{a=2}_{1/2} \ ) Y_{e} \nonumber \\
&+& ( \ 3 {\rm tr}[ Y_{d}^{\dagger} Y_{d} ] + {\rm tr}[ Y_{e}^{\dagger} Y_{e} ] 
- \frac{9}{5} g_{1}^{2} - 3 g_{2}^{2} \ ) A_{e} \nonumber \\
&+& A_{e} Y_{D}^{\dagger} Y_{D} + 2 Y_{e} Y_{D}^{\dagger} A_{D} \ ,
\end{eqnarray}
 where neutrino Dirac Yukawa coupling $Y_{D}$ and the corresponding A-term $A_{D}$ 
 appear when we introduce singlet neutrinos lighter than $M_{cut}$.
From (53-61), we obtain the following equations for $A_{u}$, $A_{d}$, $A_{e}$
 respectively in $Y_{u}$, $Y_{d}$, $Y_{e}$-bases:
\begin{eqnarray}
16 \pi^{2} \mu \frac{{\rm d}}{{\rm d}\mu} (U_{U} A_{u} U_{Qu}) &=& 
3 U_{U} A_{u} Y_{u}^{\dagger} Y_{u} U_{Qu} + 3 U_{U} Y_{u} Y_{u}^{\dagger} A_{u} U_{Qu} \nonumber \\
&+& (U_{U} A_{u} U_{Qu}) ({\rm diagonal \ part \ of} \ U_{Qu}^{\dagger} Y_{d}^{\dagger} Y_{d} U_{Qu}) + 2 U_{U} Y_{u} Y_{d}^{\dagger} A_{d} U_{Qu} \nonumber \\
&+& 2 ( \ 3 {\rm tr}[ Y_{u}^{\dagger} A_{u} ] 
- \frac{13}{15} g_{1}^{2} M^{a=1}_{1/2} - 3 g_{2}^{2} M^{a=2}_{1/2} - \frac{16}{3} g_{3}^{2} M^{a=3}_{1/2} \ ) (U_{U} Y_{u} U_{Qu}) \nonumber \\
&+& ( \ 3 {\rm tr}[ Y_{u}^{\dagger} Y_{u} ] - \frac{13}{15} g_{1}^{2} - 3 g_{2}^{2} - \frac{16}{3} g_{3}^{2} \ ) (U_{U} A_{u} U_{Qu}) \nonumber \\
&+& {\rm tr}[ Y_{D}^{\dagger} Y_{D} ] (U_{U} A_{u} U_{Qu}) + {\rm tr}[ Y_{D}^{\dagger} A_{D} ] (U_{U} Y_{u} U_{Qu}) \ ,
\\
16 \pi^{2} \mu \frac{{\rm d}}{{\rm d}\mu} (U_{D} A_{d} U_{Qd}) &=& 
3 U_{D} A_{d} Y_{d}^{\dagger} Y_{d} U_{Qd} + 3 U_{D} Y_{d} Y_{d}^{\dagger} A_{d} U_{Qd} \nonumber \\
&+& (U_{D} A_{d} U_{Qd}) ({\rm diagonal \ part \ of} \ U_{Qd}^{\dagger} Y_{u}^{\dagger} Y_{u} U_{Qd}) + 2 U_{D} Y_{d} Y_{u}^{\dagger} A_{u} U_{Qd} \nonumber \\
&+& 2 ( \ 3 {\rm tr}[ Y_{d}^{\dagger} A_{d} ] + {\rm tr}[ Y_{e}^{\dagger} A_{e} ]
- \frac{7}{15} g_{1}^{2} M^{a=1}_{1/2} - 3 g_{2}^{2} M^{a=2}_{1/2} - \frac{16}{3} g_{3}^{2} M^{a=3}_{1/2} \ ) (U_{D} Y_{d} U_{Qd})\nonumber \\
&+& ( \ 3 {\rm tr}[ Y_{d}^{\dagger} Y_{d} ] + {\rm tr}[ Y_{e}^{\dagger} Y_{e} ] 
- \frac{7}{15} g_{1}^{2} - 3 g_{2}^{2} - \frac{16}{3} g_{3}^{2} \ ) (U_{D} A_{d} U_{Qd}) \ ,
\\
16 \pi^{2} \mu \frac{{\rm d}}{{\rm d}\mu} (U_{E} A_{e} U_{L}) &=& 
3 U_{E} A_{e} Y_{e}^{\dagger} Y_{e} U_{L} + 3 U_{E} Y_{e} Y_{e}^{\dagger} A_{e} U_{L} \nonumber \\
&+& 2 ( \ 3 {\rm tr}[ Y_{d}^{\dagger} A_{d} ] + {\rm tr}[ Y_{e}^{\dagger} A_{e} ]
- \frac{9}{5} g_{1}^{2} M^{a=1}_{1/2} - 3 g_{2}^{2} M^{a=2}_{1/2} \ ) (U_{E} Y_{e} U_{L}) \nonumber \\
&+& ( \ 3 {\rm tr}[ Y_{d}^{\dagger} Y_{d} ] + {\rm tr}[ Y_{e}^{\dagger} Y_{e} ] 
- \frac{9}{5} g_{1}^{2} - 3 g_{2}^{2} \ ) (U_{E} A_{e} U_{L})\nonumber \\
&+& (U_{E} A_{e} U_{L}) ({\rm diagonal \ part \ of} \ U_{L}^{\dagger} Y_{D}^{\dagger} Y_{D} U_{L}) + 2 U_{E} Y_{e} Y_{D}^{\dagger} A_{D} U_{L} \ ,
\end{eqnarray}
To study the effects of MFV, we set the initial conditions 
 for $A_{u}$, $A_{d}$, $A_{e}$ as 
\begin{eqnarray*}
(A_{u})_{ij}\vert_{{\rm ini.}} &=& M_{u} (Y_{u})_{ij} \ , \ \ \ \ \ 
(A_{d})_{ij}\vert_{{\rm ini.}} \ = \ M_{d} (Y_{d})_{ij} \ , \ \ \ \ \ 
(A_{e})_{ij}\vert_{{\rm ini.}} \ = \ M_{e} (Y_{e})_{ij}  
\end{eqnarray*}
with mass parameters, $M_{u}$, $M_{d}$ and $M_{e}$. 
Then the terms $2 U_{U} Y_{u} Y_{d}^{\dagger} A_{d} U_{Qu}$ in (62),
$2 U_{D} Y_{d} Y_{u}^{\dagger} A_{u} U_{Qd}$ in (63)
and $2 U_{E} Y_{e} Y_{D}^{\dagger} A_{D} U_{L}$ in (64)
respectively give rise to 
off-diagonal terms of $(U_{U} A_{u} U_{Qu})$, $(U_{D} A_{d} U_{Qd})$, $(U_{E} A_{e} U_{L})$,
which were initially diagonal.
These off-diagonal terms in turn generate off-diagonal terms through (62-64),
but this does not change the orders of themselves.
Noting that the orders of the Yukawa components in each basis are given as
($\delta_{ij}$ is the ordinary Kronecker's delta):
\begin{eqnarray*}
(U_{U} Y_{u} U_{Qu})_{ij} &\sim& \beta_{i}\alpha_{i} \ \delta_{ij} \ , \ \ \ \ \ 
(U_{U} Y_{d} U_{Qu})_{ij} \ \sim \ \gamma_{i}\alpha_{j} \ , \\
(U_{D} Y_{u} U_{Qd})_{ij} &\sim& \beta_{i}\alpha_{j} \ , \ \ \ \ \ 
(U_{D} Y_{d} U_{Qd})_{ij} \ \sim \ \gamma_{i}\alpha_{i} \ \delta_{ij} \ , \\
(U_{E} Y_{e} U_{L})_{ij} &\sim& \epsilon_{i}\delta_{i} \ \delta_{ij} \ , \ \ \ \ \
(U_{E} Y_{D} U_{L})_{ij} \ \sim \ \zeta_{i}\delta_{j} \ ,
\end{eqnarray*}
where $\zeta_{i}$'s are the geometrical factors for singlet neutrinos
satisfying $\zeta_{i} \lesssim 1$,
we estimate the orders of the off-diagonal terms of 
$(U_{U} A_{u} U_{Qu})$, $(U_{D} A_{d} U_{Qd})$, $(U_{E} A_{e} U_{L})$
that arise through RG running as ($i \neq j$):
\begin{eqnarray}
\Delta (U_{U} A_{u} U_{Qu})_{ij} &\sim&
\frac{1}{16 \pi^{2}} \ \ln(\frac{M_{cut}}{M_{W}}) \ 
2 (U_{U} Y_{u} Y_{d}^{\dagger} A_{d} U_{Qu})_{ij} \nonumber \\
&\sim& \frac{2}{16 \pi^{2}} \ \ln(\frac{M_{cut}}{M_{W}}) \ 
\beta_{i}\alpha_{i} \ \alpha_{i}(\gamma_{3})^{2}\alpha_{j} \ M_{u} \ , \\
\Delta (U_{D} A_{d} U_{Qd})_{ij} &\sim&
\frac{1}{16 \pi^{2}} \ \ln(\frac{M_{cut}}{M_{W}}) \ 
2 (U_{D} Y_{d} Y_{u}^{\dagger} A_{u} U_{Qd})_{ij} \nonumber \\
&\sim& \frac{2}{16 \pi^{2}} \ \ln(\frac{M_{cut}}{M_{W}}) \
\gamma_{i}\alpha_{i} \ \alpha_{i}(\beta_{3})^{2}\alpha_{j} \ M_{d} \ , \\
\Delta (U_{E} A_{e} U_{L})_{ij} &\sim&
\frac{1}{16 \pi^{2}} \ \ln(\frac{M_{cut}}{M_{seesaw}}) \ 
2 (U_{E} Y_{e} Y_{D}^{\dagger} A_{D} U_{L})_{ij} \nonumber \\
&\sim& \frac{2}{16 \pi^{2}} \ \ln(\frac{M_{cut}}{M_{seesaw}}) \
\epsilon_{i}\delta_{i} \ \delta_{i}(\zeta_{3})^{2}\delta_{j} \ M_{e} \ ,
\end{eqnarray}
where $M_{seesaw}$ indicates the mass scale of the singlet neutrinos if they exist.
We used the approximation that $\sum_{k} (\gamma_{k})^{2} \simeq (\gamma_{3})^{2}$,
$\sum_{k} (\beta_{k})^{2} \simeq (\beta_{3})^{2}$ and 
$\sum_{k} (\zeta_{k})^{2} \simeq (\zeta_{3})^{2}$.
As for diagonal terms of  
$(U_{U} A_{u} U_{Qu})$, $(U_{D} A_{d} U_{Qd})$, $(U_{E} A_{e} U_{L})$,
the equations (65-67) do not change their orders.
In conclusion, the orders of MFV effects on A-terms
are given by the estimates (65-67).

We next estimate the orders of MFV effects on matter soft mass terms,
$m_{Q}^{2}$, $m_{U}^{2}$, $m_{D}^{2}$, $m_{L}^{2}$, $m_{E}^{2}$,
in the basis where $Y_{u}$ or $Y_{d}$ and $Y_{e}$ are diagonal.
Below is the list of those terms in MSSM RG equations that give rise to
flavor non-universal soft masses:
\begin{eqnarray}
16 \pi^{2} \mu \frac{{\rm d}}{{\rm d}\mu} m_{Q}^{2} &\supset& 
 Y_{u}^{\dagger} Y_{u} m_{Q}^{2} + m_{Q}^{2} Y_{u}^{\dagger} Y_{u}
 + 2 Y_{u}^{\dagger} m_{U}^{2} Y_{u} + 2 (Y_{u}^{\dagger} Y_{u}) m_{H_{u}}^{2} \nonumber \\
&+& Y_{d}^{\dagger} Y_{d} m_{Q}^{2} + m_{Q}^{2} Y_{d}^{\dagger} Y_{d}
 + 2 Y_{d}^{\dagger} m_{D}^{2} Y_{d} + 2 (Y_{d}^{\dagger} Y_{d}) m_{H_{d}}^{2} \nonumber \\
&+& 2 A_{u}^{\dagger}A_{u} + 2 A_{d}^{\dagger}A_{d} \ ,
\\
16 \pi^{2} \mu \frac{{\rm d}}{{\rm d}\mu} m_{U}^{2} &\supset& 
2 Y_{u} Y_{u}^{\dagger} m_{U}^{2} + 2 m_{U}^{2} Y_{u} Y_{u}^{\dagger}
 + 4 Y_{u} m_{Q}^{2} Y_{u}^{\dagger} + 4 (Y_{u} Y_{u}^{\dagger}) m_{H_{u}}^{2} \nonumber \\
&+& 4 A_{u}A_{u}^{\dagger} \ ,
\\ 
16 \pi^{2} \mu \frac{{\rm d}}{{\rm d}\mu} m_{D}^{2} &\supset& 
2 Y_{d} Y_{d}^{\dagger} m_{D}^{2} + 2 m_{D}^{2} Y_{d} Y_{d}^{\dagger}
 + 4 Y_{d} m_{Q}^{2} Y_{d}^{\dagger} + 4 (Y_{d} Y_{d}^{\dagger}) m_{H_{d}}^{2} \nonumber \\
&+& 4 A_{d}A_{d}^{\dagger} \ ,
\\ 16 \pi^{2} \mu \frac{{\rm d}}{{\rm d}\mu} m_{L}^{2} &\supset& 
Y_{e}^{\dagger} Y_{e} m_{L}^{2} + m_{L}^{2} Y_{e}^{\dagger} Y_{e}
 + 2 Y_{e}^{\dagger} m_{E}^{2} Y_{e} + 2 (Y_{e}^{\dagger} Y_{e}) m_{H_{d}}^{2} \nonumber \\
&+& 2 A_{e}^{\dagger}A_{e} \nonumber \\
&+& Y_{D}^{\dagger} Y_{D} m_{L}^{2} + m_{L}^{2} Y_{D}^{\dagger} Y_{D}
+ 2 Y_{D}^{\dagger} m_{N}^{2} Y_{D} 
+ 2 (Y_{D}^{\dagger} Y_{D}) m_{H_{u}}^{2} + 2 A_{D}^{\dagger}A_{D} \ ,
\\
16 \pi^{2} \mu \frac{{\rm d}}{{\rm d}\mu} m_{E}^{2} &\supset& 
2 Y_{e} Y_{e}^{\dagger} m_{E}^{2} + 2 m_{E}^{2} Y_{e} Y_{e}^{\dagger}
 + 4 Y_{e} m_{L}^{2} Y_{e}^{\dagger} + 4 (Y_{e} Y_{e}^{\dagger}) m_{H_{d}}^{2} \nonumber \\
&+& 4 A_{e}A_{e}^{\dagger} \ ,
\end{eqnarray}
where again, $Y_{D}$ and $A_{D}$ appear when singlet neutrinos lighter than $M_{cut}$ exist.
 From (53-58) and (68-72), we obtain the following equations for 
$m_{Q}^{2}$ in $Y_{u}$-diagonal basis, $m_{U}^{2}$ in $Y_{u}$-diagonal basis,
$m_{D}^{2}$ in $Y_{d}$-diagonal basis, $m_{L}^{2}$ in $Y_{e}$-diagonal basis
and $m_{E}^{2}$ in $Y_{e}$-diagonal basis:
\begin{eqnarray}
16 \pi^{2} \mu \frac{{\rm d}}{{\rm d}\mu} (U_{Qu}^{\dagger} m_{Q}^{2} U_{Qu}) &\supset& 
U_{Qu}^{\dagger} Y_{u}^{\dagger} Y_{u} m_{Q}^{2} U_{Qu} + U_{Qu}^{\dagger} m_{Q}^{2} Y_{u}^{\dagger} Y_{u} U_{Qu} \nonumber \\
&+& 2 U_{Qu}^{\dagger} Y_{u}^{\dagger} m_{U}^{2} Y_{u} U_{Qu} + 2 (U_{Qu}^{\dagger} Y_{u}^{\dagger} Y_{u} U_{Qu}) m_{H_{u}}^{2} \nonumber \\
&+& U_{Qu}^{\dagger} ({\rm diagonal \ parts \ of} \ Y_{d}^{\dagger} Y_{d}) U_{Qu} (U_{Qu}^{\dagger} m_{Q}^{2} U_{Qu}) \nonumber \\
&+& (U_{Qu}^{\dagger} m_{Q}^{2} U_{Qu}) U_{Qu}^{\dagger} ({\rm diagonal \ parts \ of} \ Y_{d}^{\dagger} Y_{d}) U_{Qu} \nonumber \\
&+& 2 U_{Qu}^{\dagger} Y_{d}^{\dagger} m_{D}^{2} Y_{d} U_{Qu} + 2 (U_{Qu}^{\dagger} Y_{d}^{\dagger} Y_{d} U_{Qu}) m_{H_{d}}^{2} \nonumber \\
&+& 2 U_{Qu}^{\dagger}A_{u}^{\dagger}A_{u}U_{Qu} + 2 U_{Qu}^{\dagger}A_{d}^{\dagger}A_{d}U_{Qu} \ ,
\\
16 \pi^{2} \mu \frac{{\rm d}}{{\rm d}\mu} (U_{U} m_{U}^{2} U_{U}^{\dagger}) &\supset& 
2 U_{U} Y_{u} Y_{u}^{\dagger} m_{U}^{2} U_{U}^{\dagger} + 2 U_{U} m_{U}^{2} Y_{u} Y_{u}^{\dagger} U_{U}^{\dagger} \nonumber \\
&+& 4 U_{U} Y_{u} m_{Q}^{2} Y_{u}^{\dagger} U_{U}^{\dagger} + 4 (U_{U} Y_{u} Y_{u}^{\dagger} U_{U}^{\dagger}) m_{H_{u}}^{2} \nonumber \\
&+& 4 U_{U}A_{u}A_{u}^{\dagger}U_{U}^{\dagger} \ ,
\\ 
16 \pi^{2} \mu \frac{{\rm d}}{{\rm d}\mu} (U_{D} m_{D}^{2} U_{D}^{\dagger}) &\supset& 
2 U_{D} Y_{d} Y_{d}^{\dagger} m_{D}^{2} U_{D}^{\dagger} + 2 U_{D} m_{D}^{2} Y_{d} Y_{d}^{\dagger} U_{D}^{\dagger} \nonumber \\
&+& 4 U_{D} Y_{d} m_{Q}^{2} Y_{d}^{\dagger} U_{D}^{\dagger} + 4 (U_{D} Y_{d} Y_{d}^{\dagger} U_{D}^{\dagger}) m_{H_{d}}^{2} \nonumber \\
&+& 4 U_{D}A_{d}A_{d}^{\dagger}U_{D}^{\dagger} \ ,
\\ 16 \pi^{2} \mu \frac{{\rm d}}{{\rm d}\mu} (U_{L}^{\dagger} m_{L}^{2} U_{L}) &\supset& 
U_{L}^{\dagger} Y_{e}^{\dagger} Y_{e} m_{L}^{2} U_{L} + U_{L}^{\dagger} m_{L}^{2} Y_{e}^{\dagger} Y_{e} U_{L} \nonumber \\
&+& 2 U_{L}^{\dagger} Y_{e}^{\dagger} m_{E}^{2} Y_{e} U_{L} + 2 (U_{L}^{\dagger} Y_{e}^{\dagger} Y_{e} U_{L}) m_{H_{d}}^{2} \nonumber \\
&+& 2 U_{L}^{\dagger}A_{e}^{\dagger}A_{e}U_{L} \nonumber \\
&+& U_{L}^{\dagger} ({\rm diagonal \ parts \ of} \ Y_{D}^{\dagger} Y_{D}) U_{L} (U_{L}^{\dagger} m_{L}^{2} U_{L}) \nonumber \\
&+& (U_{L}^{\dagger} m_{L}^{2} U_{L}) U_{L}^{\dagger} ({\rm diagonal \ parts \ of} \ Y_{D}^{\dagger} Y_{D}) U_{L} \nonumber \\
&+& 2 U_{L}^{\dagger} Y_{D}^{\dagger} m_{N}^{2} Y_{D} U_{L}
+ 2 (U_{L}^{\dagger} Y_{D}^{\dagger} Y_{D} U_{L}) m_{H_{u}}^{2} + 2 U_{L}^{\dagger}A_{D}^{\dagger}A_{D}U_{L} \ ,
\\
16 \pi^{2} \mu \frac{{\rm d}}{{\rm d}\mu} (U_{E} m_{E}^{2} U_{E}^{\dagger}) &\supset& 
2 U_{E} Y_{e} Y_{e}^{\dagger} m_{E}^{2} U_{E}^{\dagger} + 2 U_{E} m_{E}^{2} Y_{e} Y_{e}^{\dagger} U_{E}^{\dagger} \nonumber \\
&+& 4 U_{E} Y_{e} m_{L}^{2} Y_{e}^{\dagger} U_{E}^{\dagger} + 4 (U_{E} Y_{e} Y_{e}^{\dagger} U_{E}^{\dagger}) m_{H_{d}}^{2} \nonumber \\
&+& 4 U_{E}A_{e}A_{e}^{\dagger}U_{E}^{\dagger} \ .
\end{eqnarray}
To extract the effects of MFV on matter soft mass terms,
we consider the case with the following initial conditions:
\begin{eqnarray*}
(m_{*}^{2})_{ij}\vert_{{\rm ini.}} &=& m_{*0}^{2} \delta_{ij} \ \ \ (*=Q, U, D, L, E, N) \ , \\
(A_{u})_{ij}\vert_{{\rm ini.}} &=& M_{u} (Y_{u})_{ij} \ , \ \ \ \ \ 
(A_{d})_{ij}\vert_{{\rm ini.}} \ = \ M_{d} (Y_{d})_{ij} \ , \ \ \ \ \ 
(A_{e})_{ij}\vert_{{\rm ini.}} \ = \ M_{e} (Y_{e})_{ij} \ , \\
(A_{D})_{ij}\vert_{{\rm ini.}} &=& M_{D} (Y_{D})_{ij} \ .
\end{eqnarray*}
In (71), the terms involving $Y_{d}$ or $A_{d}$ induce the off-diagonal terms of 
$(U_{Qu}^{\dagger} m_{Q}^{2} U_{Qu})$ of the order ($i \neq j$):
\begin{eqnarray}
\Delta (U_{Qu}^{\dagger} m_{Q}^{2} U_{Qu})_{ij} &\sim&
\frac{1}{16 \pi^{2}} \ \ln( \frac{M_{cut}}{M_{W}} ) \
 \alpha_{i} (\gamma_{3})^{2} \alpha_{j} \ 
 \{ \ 2 m_{Q0}^{2} + 2 m_{D0}^{2} + 2 m_{Hd}^{2} + 2 M_{d}^{2} \ \} \ .
\end{eqnarray}
It is then clear that, in $Y_{d}$-diagonal basis,
$(U_{Qd}^{\dagger} m_{Q}^{2} U_{Qd})$ obtain the off-diagonal terms of the order ($i \neq j$):
\begin{eqnarray}
\Delta (U_{Qd}^{\dagger} m_{Q}^{2} U_{Qd})_{ij} &\sim&
\frac{1}{16 \pi^{2}} \ \ln( \frac{M_{cut}}{M_{W}} ) \
 \alpha_{i} (\beta_{3})^{2} \alpha_{j} \ 
 \{ \ 2 m_{Q0}^{2} + 2 m_{U0}^{2} + 2 m_{Hu}^{2} + 2 M_{u}^{2} \ \} \ .
\end{eqnarray}
If singlet neutrinos lighter than $M_{cut}$ exist,
the terms in (74) involving $Y_{D}$ or $A_{D}$ induce the off-diagonal terms of 
$(U_{L}^{\dagger} m_{L}^{2} U_{L})$ of the order ($i \neq j$):
\begin{eqnarray}
\Delta (U_{L}^{\dagger} m_{L}^{2} U_{L})_{ij} &\sim&
\frac{1}{16 \pi^{2}} \ \ln( \frac{M_{cut}}{M_{seesaw}} ) \
 \delta_{i} (\zeta_{3})^{2} \delta_{j} \ 
 \{ \ 2 m_{L0}^{2} + 2 m_{N0}^{2} + 2 m_{Hu}^{2} + 2 M_{D}^{2} \ \} \ .
\end{eqnarray}
On the other hand, for $(U_{U} m_{U}^{2} U_{U}^{\dagger})$,
off-diagonal terms arise from the following two terms in (74):
\begin{eqnarray}
4 U_{U} Y_{u} m_{Q}^{2} Y_{u}^{\dagger} U_{U}^{\dagger} &=& 4 (U_{U} Y_{u} U_{Qu}^{\dagger})
(U_{Qu} m_{Q}^{2} U_{Qu}^{\dagger}) (U_{Qu}^{\dagger} Y_{u}^{\dagger} U_{U}^{\dagger}) \ , \nonumber
\\
4U_{U}A_{u}A_{u}^{\dagger}U_{U}^{\dagger} &=& 
4 (U_{U} A_{u} U_{Qu}^{\dagger})(U_{Qu}^{\dagger} A_{u}^{\dagger} U_{U}^{\dagger}) \ .
\end{eqnarray}
Off-diagonal terms of $(U_{Qu}^{\dagger} m_{Q}^{2} U_{Qu})$ induced through (73)
and those of $(U_{U}A_{u}U_{Qu})$ induced through (62)
in turn give rise to off-diagonal terms of $(U_{U} m_{U}^{2} U_{U}^{\dagger})$ through (81).
Therefore, from (78) and (65), we estimate off-diagonal terms of
$(U_{U} m_{U}^{2} U_{U}^{\dagger})$  induced by RG running as ($i \neq j$):
\begin{eqnarray}
\Delta(U_{U} m_{U}^{2} U_{U}^{\dagger})_{ij} &\sim& 
\left[ \frac{1}{16 \pi^{2}} \ln(\frac{M_{cut}}{M_{W}}) \right]^{2} \nonumber \\
&\times& [ \ 4 \beta_{i}\alpha_{i} \alpha_{i} (\gamma_{3})^{2} \alpha_{j} \alpha_{j}\beta_{j} 
( 2 m_{Q0}^{2} + 2 m_{D0}^{2} + 2 m_{Hd}^{2} + 2 M_{d}^{2} ) \nonumber \\
&+& 4 \sum_{k} \ ( \ \beta_{i}\alpha_{i} \delta_{ik} M_{u} + \beta_{i}\alpha_{i} \alpha_{i} (\gamma_{3})^{2} \alpha_{k} M_{u} \ )
( \ \alpha_{k}\beta_{k} \delta_{kj} M_{u} + \alpha_{k} (\gamma_{3})^{2} \alpha_{j} \alpha_{j} \beta_{j} \ ) \ ] \nonumber \\
&\sim& \left[ \frac{1}{16 \pi^{2}} \ln(\frac{M_{cut}}{M_{W}}) \right]^{2} \nonumber \\
&\times& [ \ 4 \beta_{i} (\alpha_{i})^{2} (\gamma_{3})^{2} (\alpha_{j})^{2} \beta_{j} \ 
( 2 m_{Q0}^{2} + 2 m_{D0}^{2} + 2 m_{Hd}^{2} + 2 M_{d}^{2} ) \nonumber \\
&+& 8 \beta_{i} (\alpha_{i})^{2} (\gamma_{3})^{2} (\alpha_{j})^{2} \beta_{j} M_{u}^{2}
\ + \ 4 \beta_{i} (\alpha_{i})^{2} (\gamma_{3})^{2} (\alpha_{3})^{2} (\gamma_{3})^{2} (\alpha_{j})^{2} \beta_{j} M_{u}^{2} \ ] \ .
\end{eqnarray}
Likewise, we obtain the following estimates on off-diagonal terms of 
$(U_{D} m_{D}^{2} U_{D}^{\dagger})$ and $(U_{E} m_{E}^{2} U_{E}^{\dagger})$ ($i \neq j$):
\begin{eqnarray}
\Delta(U_{D} m_{D}^{2} U_{D}^{\dagger})_{ij}
&\sim& \left[ \frac{1}{16 \pi^{2}} \ln(\frac{M_{cut}}{M_{W}}) \right]^{2} \nonumber \\
&\times& [ \ 4 \gamma_{i} (\alpha_{i})^{2} (\beta_{3})^{2} (\alpha_{j})^{2} \gamma_{j} \ 
( 2 m_{Q0}^{2} + 2 m_{U0}^{2} + 2 m_{Hu}^{2} + 2 M_{u}^{2} ) \nonumber \\
&+& 8 \gamma_{i} (\alpha_{i})^{2} (\beta_{3})^{2} (\alpha_{j})^{2} \gamma_{j} M_{d}^{2}
\ + \ 4 \gamma_{i} (\alpha_{i})^{2} (\beta_{3})^{2} (\alpha_{3})^{2} (\beta_{3})^{2} (\alpha_{j})^{2} \gamma_{j} M_{d}^{2} \ ] \ , \\
\Delta(U_{E} m_{E}^{2} U_{E}^{\dagger})_{ij}
&\sim& \left[ \frac{1}{16 \pi^{2}} \ln(\frac{M_{cut}}{M_{seesaw}}) \right]^{2} \nonumber \\
&\times& [ \ 4 \epsilon_{i} (\delta_{i})^{2} (\zeta_{3})^{2} (\delta_{j})^{2} \epsilon_{j} \ 
( 2 m_{L0}^{2} + 2 m_{N0}^{2} + 2 m_{Hu}^{2} + 2 M_{D}^{2} ) \nonumber \\
&+& 8 \epsilon_{i} (\delta_{i})^{2} (\zeta_{3})^{2} (\delta_{j})^{2} \epsilon_{j} M_{d}^{2}
\ + \ 4 \epsilon_{i} (\delta_{i})^{2} (\zeta_{3})^{2} (\delta_{3})^{2} (\zeta_{3})^{2} (\delta_{j})^{2} \epsilon_{j} M_{e}^{2} \ ] \ .
\end{eqnarray}
In summary, the orders of MFV effects on soft mass terms 
 are given in (79, 80), (82-84).

We have estimated the orders of MFV effects on A-terms and soft mass terms
 in the basis where $Y_{u}$ or $Y_{d}$ and $Y_{e}$ are diagonal.
In the rest of the section, we compare MFV effects 
 with flavor-violating gravity mediation effects of our model
 and discuss their difference.

Flavor-violating gravity mediation effects at the scale $M_{cut}$ 
 can be read from (46-49).
We assume that the couplings in 5D theory, $d_{*}$, $a_{*}$, are all $O(1)$.
This is a natural assumption because we are trying to explain 
 the hierarchy of 4D theory from 5D geometrical point of view.
In an arbitrary basis, the flavor-violating parts of 
 A-terms that arise from gravity mediation 
 and are not proportional to the corresponding Yukawa couplings 
 are given by
\begin{eqnarray}
(A_{u})_{ij} &\supset& \ \sim \ \beta_{i}\alpha_{j} M_{X} \ , \\
(A_{d})_{ij} &\supset& \ \sim \ \gamma_{i}\alpha_{j} M_{X} \ , \\
(A_{e})_{ij} &\supset& \ \sim \ \epsilon_{i}\delta_{j} M_{X}
\end{eqnarray}
at the scale $M_{cut}$,
where $M_{X}$ is defined as
\begin{eqnarray*}
M_{X} &\equiv& \frac{\vert < F_{\tilde{X}} > \vert}{M_{5} e^{-kR\pi}} \ .
\end{eqnarray*}
Matter soft mass terms that arise from gravity mediation are given by
\begin{eqnarray}
(m_{Q}^{2})_{ij} &\sim& \alpha_{i}\alpha_{j} M_{X}^{2} \ , \nonumber \\
{\rm (Q, \ \alpha)} \ & \rightarrow & \ {\rm (U, \beta), \ (D, \gamma), \ (L, \delta), \ (E, \epsilon)}
\end{eqnarray}
at the scale $M_{cut}$.
First, we argue that, at the electroweak scale,
the flavor-violating parts of A-terms 
 that are not proportional to the corresponding Yukawa couplings
 are still estimated as in (85-87)
 and flavor-violating parts of matter soft mass terms are estimated as in (88).
This is understood from the form of RG equations;
the right hand sides of the RG equations (59-61) 
 depend on the Yukawa couplings and A-terms themselves.
For the component $(A_{u})_{ij}$, the right hand side of (59) 
 is at least proportional to $\beta_{i}\alpha_{j}$. 
Similarly, the right hand sides of (68-72), 
 which express flavor-violating contributions,
 depend on the Yukawa couplings and A-terms. 
The flavor-violating part of the RG equation for the component $(m_{Q}^{2})_{ij}$ 
 is at least proportional to $\alpha_{i}\alpha_{j}$.
The same discussion applies to other A-terms and matter soft masses,
 and we conclude that RG running keeps
 the orders of flavor-violating parts of A-terms 
 and matter soft mass terms as in (85-88).
Second, we argue that the estimates (85-88) are valid 
 even in $Y_{u}$ or $Y_{d}$ and $Y_{e}$-diagonal basis.
This is because the 5D couplings $(y_{*})_{ij}$,
 $(a_{*})_{ij}$, $d_{*1ij}$ and $d_{*2ij}$ in (46-49)
 are independent of each other. 
Therefore the matrices $(a_{*})_{ij}$, $d_{*1ij}$ and $d_{*2ij}$, 
 which give rise to A-terms and matter soft masses, 
 are arbitrary even when $(y_{*})_{ij}$ is diagonal.

Let us compare the orders of MFV effects (65-67, 79, 80, 82-84)
and those of flavor-violating gravity mediation effects (85-88).
We assume that $M_{u}, M_{d}, M_{e}$ in (65-67) and $m_{*0}^{2}, m_{Hu}^{2}, m_{Hd}^{2}$ in (79, 80, 82-84)
are of the same order as $M_{X}$ in (85-88).
For the A-term components $(A_{u})_{1j}$, $(A_{u})_{2j}$ 
 in $Y_{u}$-diagonal basis and $(A_{d})_{1j}$, $(A_{d})_{2j}$ 
 in $Y_{d}$-diagonal basis,
 MFV effects are always much smaller than flavor-violating gravity mediation effects
because the former are suppressed by $(\alpha_{1})^{2}$ or $(\alpha_{2})^{2}$
compared to the latter. 
For the components $(A_{u})_{3j}, (A_{d})_{3j}$,
MFV effects can be of the same order as flavor-violating gravity mediation effects.
For the A-term $(A_{e})_{ij}$, MFV effects are much smaller than flavor-violating gravity mediation effects 
when the order of $\delta_{3}$ is significantly smaller than $1$ 
 (we will see in the next section that 
 this is the case for a realistic mass spectrum).
If singlet neutrinos lighter than $M_{cut}$ do \textit{not} exist,
 $A_{e}$ is diagonal.

For the soft mass terms $m_{Q}^{2}$ and $m_{L}^{2}$ (if singlet neutrinos lighter than $M_{cut}$ exist),
MFV effects can be of the same order as flavor-violating gravity mediation effects.
For the terms $m_{U}^{2}$ in $Y_{u}$-diagonal basis and $m_{D}^{2}$ in $Y_{d}$-diagonal basis,
MFV effects are always much smaller than flavor-violating gravity mediation effects 
except their (3,3)-components.
This is because the components of these terms other than (3,3) 
are at least suppressed by $(\alpha_{1})^{2}$ or $(\alpha_{2})^{2}$.
For the components $(m_{U}^{2})_{33}$ in $Y_{u}$-diagonal basis and $(m_{D}^{2})_{33}$ in $Y_{d}$-diagonal basis,
the former can be as large as the latter.
For the term $m_{E}^{2}$ in $Y_{e}$-diagonal basis,
MFV effects are much smaller than flavor-violating gravity mediation effects 
when the order of $\delta_{3}$ is significantly smaller than $1$.
If singlet neutrinos lighter than $M_{cut}$ do \textit{not} exist,
there is no MFV on $m_{L}^{2}$, $m_{E}^{2}$.

In this section, we have discussed the difference 
 between the flavor-violating soft terms in MFV 
 and those generated by the gravity mediation of our model.
We have proved that, for some components of A-terms and soft mass terms,
 the gravity mediation contribution dominates. 
Therefore, it is in principle possible to distinguish our model 
 from other SUSY models with MFV.

\section{Particle mass spectra and experimental constraints}

We calculate a sample of mass spectra and check 
 that this model provides a realistic mass spectrum 
 consistent with current experimental bounds.

Our numerical analysis is done in the following way. 
We fix the cutoff scale, $M_{cut}$, which is of the same order 
 as the KK scale, $M_{5} e^{-kR\pi}$, from the relation (41)
\begin{eqnarray*}
M_{cut} \ \sim \ \delta_{3}^{2} \ 
 \frac{v_{u}^{2}}{ 3 \times 10^{-11} \ {\rm GeV} } 
 \ \simeq \ \delta_{3}^{2} \times 10^{15} \ {\rm GeV} \ .
\end{eqnarray*}
We assume that contact term couplings between the MSSM fields 
 and the SUSY breaking field in (42-50) are all $O(1)$  
 and adopt the following initial condition:
\begin{eqnarray}
M_{1/2}^{a} &=& 2 \ M_{X} \ , \\
m_{H_{u}}^{2} \ = \ m_{H_{d}}^{2} &=& M_{X}^{2} \ , \\
(m_{Q}^{2})_{ij} &=& c_{Qij} \ \alpha_{i}\alpha_{j} M_{X}^{2} \ , \nonumber \\
{\rm (Q, \ \alpha)} \ & \rightarrow & \ {\rm (U, \beta), \ (D, \gamma), \ (L, \delta), \ (E, \epsilon)} \ , \\
A_{uij} &=& -M_{X} (Y_{u})_{ij} \ + \ a_{uij} \ \beta_{i} \alpha_{j} M_{X} \ , \\
A_{dij} &=& -M_{X} (Y_{d})_{ij} \ + \ a_{dij} \ \gamma_{i} \alpha_{j} M_{X} \ , \\
A_{eij} &=& -M_{X} (Y_{e})_{ij} \ + \ a_{eij} \ \epsilon_{i} \delta_{j} M_{X} \ ,
\end{eqnarray}
where $M_{X}$ was defined as 
 $M_{X} \equiv \vert <F_{\tilde{X}}> \vert / M_{5} e^{-kR\pi}$
 and we set a natural range of the parameters as 
 $0.1 \lesssim c_{*ij}, a_{*ij} \lesssim 1 $.
The factor $2$ in the right-hand side of (89) comes from 
 the factor $4(g_{4}^{a})^{2}$ in (42).
Since $M_{cut}$ is around $10^{15}$ GeV, 
 SU(2) and SU(3) couplings of MSSM, $g_{4}^{a=2}, g_{4}^{a=3}$, at $M_{cut}$
 take the value of $0.7$. 
For simplicity, we fix the normalization of U(1) coupling at 
 $M_{cut}$ as $0.7$. 
Then we obtain the factor $2$ in (89) from 
\begin{eqnarray*}
4(g_{4}^{a})^{2} [\mu_{r}=M_{cut}] &\simeq& 4 \cdot 0.7^{2} \ \simeq \ 2 \ .
\end{eqnarray*}

Our aim is to prove that, in our 5D MSSM framework, there exists a mass spectrum 
 that is consistent with 
 the current experimental bounds. 
We arrange the parameters as 
\begin{eqnarray*}
c_{*ij}, \ a_{*ij} \ = \ 1 \ \ \ {\rm for} \ i=j \ ,
\\ c_{*ij}, \ a_{*ij} \ = \ 0.1 \ \ \ {\rm for} \ i\neq j \ ,
\end{eqnarray*}
 to keep the flavor-violating terms as small as possible 
 with a mild hierarchy among the model parameters. 
Now the model has three free parameters:
\begin{eqnarray*}
M_{X}, \ \ \delta_{3} \ (\sim \delta_{2} \sim 3 \delta_{1}), \ \ \tan
 \beta \ . 
\end{eqnarray*}
The KK scale is determined by $\delta_{3}$ in (41). 
Since $\epsilon_{3}$ is smaller than ${\cal O}(1)$ from naturalness, 
 (32) leads to the condition:
\begin{eqnarray}
1 \ \gtrsim \ \delta_{3} \ \gtrsim \ m_{\tau}/ v \cos \beta 
 \ \simeq \ 0.01 / \cos \beta \ .
\end{eqnarray}

Based on this setup, we calculate mass spectra 
 for various values of $(M_{X}, \delta_{3}, \tan \beta )$
 and check if they are consistent with the current experimental
 bounds, in particular, the lower bound of Higgs boson mass. 
With the flavor-violating soft terms predicted in our model, 
 we then evaluate the rates of the lepton flavor violating processes, i.e.
 the branching ratios of $\mu \rightarrow e \gamma$ and 
 $\tau \rightarrow \mu \gamma$ decays, 
 for each sparticle mass spectrum based on the technology 
 developed in \cite{LFV} and compared the results with current bounds. 
In our analysis of MSSM RG equations, 
 we first ignore the off-diagonal terms in (91-94) from the initial
 condition and numerically solve the MSSM RG equations 
 from $M_{cut}$ to low energies using \textit{Softsusy-3.1.4} \cite{softsusy} 
 with Yukawa off-diagonal terms ignored.
After this calculation, we add the off-diagonal terms 
 to give the resultant spectrum.

Below is the list of sample values of $( M_{X}, \delta_{3}, \tan \beta )$ 
 that give realistic mass spectra consistent with the bounds 
 on Higgs boson mass and 
 $\mu \rightarrow e \gamma$ and $\tau \rightarrow \mu \gamma$ branching ratios.
We focus on the case with $M_{X} \leq 600$ GeV because 
 light mass spectra are of more phenomenological interest. 
The lightest Higgs boson mass $m_{h}$, 
 $\mu \rightarrow e \gamma$ branching ratio ($Br_{\mu}$),
 and $\tau \rightarrow \mu \gamma$ branching ratio ($Br_{\tau}$) 
 of each spectrum are also shown. 
The results shown here satisfy the current experimental bounds: 
 $m_{h} > 114.4$ GeV \cite{LEP II higgs},
 $Br_{\mu} < 1.2 \times 10^{-11}$ \cite{MEGA}
 and $Br_{\tau} < 4.5 \times 10^{-8}$ \cite{Belle}. 
The entire mass spectra for three examples are shown in the Appendix.
For larger values of $\delta_{3}$ and/or $\tan \beta$, 
 the spectrum violates the bound on $\mu \rightarrow e \gamma$ branching ratio.
For smaller $M_{X}$ and/or $\tan \beta$, 
 the lightest Higgs boson is too light.

\begin{center}
\begin{tabular}{|c||c|c||c|c|c|} \hline
$M_{X}$ (GeV)                   & 500    & 500    & 600    & 600    & 600    \\ \hline
$\tan \beta$                    & 6      & 10     & 5      & 10     & 15     \\ \hline
$\delta_{3}$                    & 0.06   & 0.1    & 0.05   & 0.1    & 0.15   \\ \hline
$m_{h}$ (GeV)                   & 115.2  & 117.5  & 114.8  & 118.4  & 119.1  \\ \hline
$Br_{\mu} \times 10^{12}$       & 7.6    & 10     & 3.9    & 4.5    & 8.2    \\ \hline
$Br_{\tau} \times 10^{12}$      & 4.3    & 6.1    & 2.1    & 2.7    & 5.7    \\ \hline
\end{tabular}
\end{center}

To $\mu \rightarrow e \gamma$ process, 
 loop diagrams containing the following terms contribute:
\begin{eqnarray*}
<H_{d}^{0}> (A_{e})_{21} &\sim& \epsilon_{2}\delta_{1} v_{d}M_{X} 
\ \sim \ (\delta_{1}/\delta_{2}) m_{\mu}M_{X} 
\ \sim \ \frac{1}{3} m_{\mu}M_{X} \ , \\
<H_{d}^{0}> (A_{e})_{12} &\sim& \epsilon_{1}\delta_{2} v_{d}M_{X}
\ \sim \ (\delta_{2}/\delta_{1}) m_{e}M_{X} 
\ \sim \ 3 m_{e}M_{X} \ , \\
(m_{L}^{2})_{12} &\sim& \delta_{1}\delta_{2}M_{X}^{2} 
\ \sim \ \frac{1}{3} (\delta_{3})^{2}M_{X}^{2} \ , \\
(m_{E}^{2})_{12} &\sim& \epsilon_{1}\epsilon_{2}M_{X}^{2} 
\ \sim \ \frac{3}{(\delta_{3})^{2}} \frac{m_{e}}{m_{\tau}} \frac{m_{\mu}}{m_{\tau}} M_{X}^{2} \ .
\end{eqnarray*}
The contributions from the terms $(A_{e})_{21}$, $(A_{e})_{12}$ 
 are almost independent of $\tan \beta$ and $\delta_{3}$. 
For the mass spectra listed above,  
 if $(A_{e})_{21}$, $(A_{e})_{12}$ were the only source of lepton flavor violation,
 they would give $Br_{\mu} \sim 7-8 \times 10^{-12}$ for $M_{X}=500$ GeV 
 and $Br_{\mu} \sim 3-4 \times 10^{-12}$ for $M_{X}=600$ GeV.
Hence we argue that, for the cases with small $\tan \beta$ and $\delta_{3}$,
 the flavor-violating A-terms give dominant contributions. 
It is obvious that $(A_{e})_{21}$ contributes 
 much more strongly than $(A_{e})_{12}$. 
On the other hand, the contribution from the term $(m_{L}^{2})_{12}$ 
 is sensitive to the values of $\tan \beta$ and $\delta_{3}$, 
 which is roughly proportional to $(\tan \beta)^{2}$ and $(\delta_{3})^{4}$.
(The net value of $Br_{\mu}$ does not reflect this rule 
 because of the interference between $(A_{e})_{21}$ contribution 
 and $(m_{L}^{2})_{12}$ contribution.)
We thus obtain the upper bounds on $\tan \beta$ and $\delta_{3}$ 
 when the contribution from the term $(m_{L}^{2})_{12}$ becomes dominant. 
The contribution from $(m_{E}^{2})_{12}$ is much suppressed 
 by the tiny ratio $m_{e}/m_{\tau}$ and has negligible impact 
 on $\mu \rightarrow e \gamma$ branching ratio.

Here we summarize the features of the sample mass spectra listed above.

(i) The typical SUSY breaking mass scale, $M_{X}$, can be as low as $500$ GeV 
and the mass spectrum is within the reach of the LHC.

(ii) The ratio $\delta_{3} / \tan \beta$ is around $0.01$,
which means that we need $\epsilon_{3} \sim 1$ to have the tau Yukawa coupling. 
Therefore the 5D superfield of singlet tau is strongly 
 localized towards the IR brane. 

(iii) $\mu \rightarrow e \gamma$ branching ratio is always higher than
O($10^{-12}$) and the model can be tested by MEG experiment \cite{MEG}.
 in the near future.  
 
(iv) $\tau \rightarrow \mu \gamma$ branching ratio is of the same order as 
$\mu \rightarrow e \gamma$ branching ratio.

The feature (ii) originates from the difference between the hierarchy of $\delta_{i}$'s
and that of $\epsilon_{i}$'s.
The experimental bound on $\mu \rightarrow e \gamma$ branching ratio
constrains the terms $(m_{L}^{2})_{12}$ and $(m_{E}^{2})_{12}$,
which are respectively proportional to $\delta_{1}\delta_{2}$ and $\epsilon_{1}\epsilon_{2}$,
with the same extent.
This gives a stronger limit on $\delta_{3}$ than on $\epsilon_{3}$
because $\delta_{i}$'s have milder hierarchy.
The orders of $\delta_{3}$ and $\epsilon_{3}$ are related through
\begin{eqnarray*}
\delta_{3}\epsilon_{3} / \tan \beta \ \sim \ m_{\tau}/v \ \simeq \ 0.01 \ ,
\end{eqnarray*}
where $\tan \beta$ cannot be smaller than about $4$ 
because otherwise the LEP II Higgs mass bound would not be satisfied.
Therefore small $\delta_{3}$ and large $\epsilon_{3}$ are favored in this model,
 which leads to the prediction that $\epsilon_{3} \sim 1$.

The feature (iii) results from the existence of 
 the flavor-violating A-term, $(A_{e})_{12}$,
 and the fact that its contribution is independent of $\tan \beta$ and $\delta_{3}$. 
The resultant branching ratio is within the future reach 
 of MEG experiment.

The feature (iv) is specific to this model because new physics models 
 normally predict $\tau \rightarrow \mu \gamma$ branching ratio 
 larger than $\mu \rightarrow e \gamma$ branching ratio
 as new physics is more likely to affect 3rd generation particles 
 than 1st and 2nd generations.
This feature is a consequence of the feature (ii). 
Since $\epsilon_{3} \sim 1$, SU(2) singlet stau obtains 
 large soft mass through gravity mediation on the IR brane 
 and becomes a few times heavier than SU(2) singlet smuon 
 and selectron if gravity mediation contributes positively 
 as is usually assumed. 
For $\tau \rightarrow \mu \gamma$ process, 
 the term $(A_{e})_{32} \sim \epsilon_{3}\delta_{2} M_{X}$ 
 always contributes.
One can compare its impact on $Br_{\tau}$ with that of $(A_{e})_{21}$ on $Br_{\mu}$
by comparing
\begin{eqnarray*}
<H_{d}^{0}> (A_{e})_{32} \ / \ m_{\tau} \ \sim \ M_{X} \ \ \ {\rm vs.} \ \ \ 
<H_{d}^{0}> (A_{e})_{21} \ / \ m_{\mu} \ \sim \ \frac{1}{3}M_{X} \ .
\end{eqnarray*}
The former is larger by the factor $3$.
However its effect is canceled by the larger mass of the singlet stau 
 propagating in the loop diagram containing $(A_{e})_{32}$.
The term $(m_{L}^{2})_{32}$ also contributes when $\tan \beta$ 
 is relatively large.
$(m_{L}^{2})_{32}$ is predicted to be about $3$ times larger than $(m_{L}^{2})_{12}$.
The contribution from $(m_{L}^{2})_{32}$ mainly comes from two types of diagrams,
 SU(2) singlet smuon propagating in one diagram and singlet stau
 propagating in the other.
However the latter is suppressed by the large stau mass, 
 which partly cancels the effects of large $(m_{L}^{2})_{32}$.
The term $(m_{E}^{2})_{23}$ contributes when $\tan \beta$ is relatively large.
However, again, its contribution is suppressed by the large mass 
 of the singlet stau propagating in the diagram.

Finally, we discuss the prediction of our model on 
 $\Delta m_{K}$ of $K^{0}-\bar{K}^{0}$ mixing 
 and $b \rightarrow s \gamma$ branching ratio, 
 based on the paper \cite{complete analysis}.

The following flavor-violating parameters predicted in our model 
 are relevant to the $K^{0}-\bar{K}^{0}$ mixing: 
\begin{eqnarray}
(m_{Q}^{2})_{12} &\sim& \alpha_{1}\alpha_{2}M_{X}^{2} \ \sim \
\lambda^{5} M_{X}^{2} \ = \ 5 \times 10^{-4} \ M_{X}^{2} \ , 
\nonumber \\
(m_{D}^{2})_{12} &\sim& \gamma_{1}\gamma_{2}M_{X}^{2} 
\ \sim \ \frac{1}{\lambda_{5}} \frac{m_{d}}{v_{d}} \frac{m_{s}}{v_{d}} M_{X}^{2}
\ \simeq \ 3 \times 10^{-5} \ \tan^{2} \beta \ M_{X}^{2} \ , 
\nonumber \\ 
<H_{d}^{0}>(A_{d})_{21} &\sim& \gamma_{2}\alpha_{1}v_{d}M_{X} 
\ \sim \ \frac{\alpha_{1}}{\alpha_{2}} m_{s}M_{X}
\ \simeq \ 4 \times 10^{-5} \ M_{X}^{2} \ \ \ {\rm for} \ M_{X}=500 \ {\rm GeV} \ , 
\nonumber \\  
<H_{d}^{0}>(A_{d})_{12} &\sim& \gamma_{1}\alpha_{2}v_{d}M_{X} 
\ \sim \ \frac{\alpha_{2}}{\alpha_{1}} m_{d}M_{X}
\ \simeq \ 5 \times 10^{-5} \ M_{X}^{2} \ \ \ {\rm for} 
\ M_{X}=500 \ {\rm GeV} \ . \nonumber 
\end{eqnarray}
Here we focus on the case with $M_{X}=500$ GeV.
The average squark mass is around the same scale. 
Comparing the above predictions with Table 1 in \cite{complete analysis}, 
 we see that, when $\tan \beta \lesssim 15$, 
 our predictions are below the limits that come 
 from the experimental bound on $\Delta m_{K}$.

For $b \rightarrow s \gamma$ process,
 our model predicts the following flavor-violating parameters:
\begin{eqnarray*}
(m_{Q}^{2})_{23} &\sim& \alpha_{2}\alpha_{3}M_{X}^{2} \ \sim \ \lambda^{2} M_{X}^{2} \ = \ 5 \times 10^{-2} \ M_{X}^{2} \ , \\
(m_{D}^{2})_{23} &\sim& \gamma_{2}\gamma_{3}M_{X}^{2} \ \sim \ \frac{1}{\lambda_{2}} \frac{m_{s}}{v_{d}} \frac{m_{b}}{v_{d}} M_{X}^{2}
\ \simeq \ 2 \times 10^{-4} \ \tan^{2} \beta \ M_{X}^{2} \ , \\
<H_{d}^{0}>(A_{d})_{32} &\sim& \gamma_{3}\alpha_{2}v_{d}M_{X} \ \sim \ \frac{\alpha_{2}}{\alpha_{3}} m_{b}M_{X}
\ \simeq \ 3 \times 10^{-4} \ M_{X}^{2} \ \ \ {\rm for} \ M_{X}=500{\rm GeV} \ , \\
<H_{d}^{0}>(A_{d})_{23} &\sim& \gamma_{2}\alpha_{3}v_{d}M_{X} \ \sim \ \frac{\alpha_{3}}{\alpha_{2}} m_{s}M_{X}
\ \simeq \ 4 \times 10^{-3} \ M_{X}^{2} \ \ \ {\rm for} \ M_{X}=500{\rm GeV} \ .
\end{eqnarray*}
Again, we focus on the case with $M_{X}=500$ GeV. 
Comparing the above predictions with Table 6 in \cite{complete analysis}.
 we find that, regardless of $\tan \beta$,
 our predictions are far below the limits that 
 come from the experimental bound adopted by \cite{complete analysis}. 
Even if we adopt the stronger bound in \cite{bsg}, 
 our predictions are still below the limits.

\section{Unusual NLSP and its flavor-violating decay}

In this section, we consider NLSP,
 which is mostly composed of SU(2) singlet charged sleptons.
As we found in the previous section,
 our model favors $\epsilon_{3} \sim 1$,
 i.e. the singlet stau superfield is localized towards the IR brane
 and receives large gravity mediation effects.
This changes the flavor structure of charged slepton mass matrix.
We here discuss the model's predictions on the flavor composition of NLSP.
 In the following, we fix $\epsilon_{3} = 1$.

We first review the charged slepton mass matrix.
Define 
\begin{eqnarray*}
 {\cal A} \equiv A_{0} - \mu \tan \beta \ ,
\end{eqnarray*}
where $A_{0}$ indicates those parts of A-terms that 
 are proportional to the corresponding Yukawa couplings.
We denote RG contributions (gaugino mediation contributions) 
 and D-term contributions to the soft masses of doublet selectron, 
 smuon, stau and singlet selectron, smuon, stau by
\begin{eqnarray*}
m_{L1}^{2}, \ m_{L2}^{2}, \ m_{L3}^{2}, \ m_{E1}^{2}, \ m_{E2}^{2}, \ m_{E3}^{2} \ .
\end{eqnarray*}
Since we are focusing on NLSP, whose candidates are singlet sleptons,
we neglect relatively small doublet slepton and doublet-singlet mixing terms.
 Further neglecting terms suppressed by $m_{e}/m_{\tau}$, we obtain the following 
 approximate form of the charged slepton mass matrix: 
\begin{eqnarray}
m_{{\rm slepton}}^{2} \sim \left(
\begin{array}{cccccc}
m_{L1}^{2}         &  0                  & 0                                      & 0          & 0                & \frac{m_{\tau} M_{X}}{3} \\
0                  &  m_{L2}^{2}         & 0                                      & 0          & {\cal A} m_{\mu} & m_{\tau} M_{X} \\
0                  &  0                  & m_{L3}^{2}                             & 0          & 0                & {\cal A} m_{\tau} + m_{\tau} M_{X} \\
0                  &  0                  & 0                                      & m_{E1}^{2} & 0                & 0 \\
0                  &  {\cal A} m_{\mu}   & 0                                      & 0          & m_{E2}^{2} + (\frac{m_{\mu}}{m_{\tau}})^{2} M_{X}^{2} & c_{\mu \tau} \ \frac{m_{\mu}}{m_{\tau}} M_{X}^{2} \\
\frac{m_{\tau} M_{X}}{3}   &  m_{\tau} M_{X}     & {\cal A} m_{\tau} + m_{\tau} M_{X}     & 0          & c_{\mu \tau} \ \frac{m_{\mu}}{m_{\tau}} M_{X}^{2} & m_{E3}^{2} + M_{X}^{2}
\end{array}
\right) \ . \nonumber \\
\end{eqnarray}
The factor $c_{\mu \tau}$ denotes the coupling of the contact term 
 among singlet smuon, singlet stau and the SUSY breaking sector, 
 and is assumed to be $O(1)$.

In the following, we consider two cases for completeness:

(i) gravity mediation contributions to soft masses are positive, 
 as is usually assumed. 

(ii) they are negative.

In case (i), SU(2) singlet stau becomes heavier 
 than singlet smuon and selectron, so that the NLSP will 
 be mainly composed of either singlet smuon or singlet selectron. 
In order to see which is the one, 
 we consider the mixing mass between singlet smuon and singlet stau, 
 namely $(m_{E}^{2})_{23}$. 
First note that the difference between $m_{E1}^{2}$ and $m_{E2}^{2}$ 
 and that between $m_{L2}^{2}$ and $m_{E2}^{2}$,
 which arise from RG running, are given by 
\begin{eqnarray}
m_{E1}^{2} - m_{E2}^{2} &\sim& \frac{16}{16 \pi^{2}} 
\ln \left( \frac{M_{cut}}{M_{W}} \right) M_{X}^{2} \ 
(y_{\mu})^{2} \tan^{2} \beta \ , \\
m_{L2}^{2} - m_{E2}^{2} &\simeq& \frac{1}{16 \pi^{2}} 
\ln \left( \frac{M_{cut}}{M_{W}} \right) \ 
\left( - \frac{18}{5} g_{1}^{2} \vert M_{1/2}^{a=1} \vert^{2} 
 + \ 6 g_{2}^{2} \vert M_{1/2}^{a=2} \vert^{2} 
+ \frac{9}{5} g_{1}^{2} S \right) \nonumber \\
&+& \{ \ -\frac{1}{2} (\cos^{2}\theta_{W} - \sin^{2}\theta_{W}) 
 M_{Z}^{2} \ \cos 2\beta 
\ + \ \sin^{2}\theta_{W} \ M_{Z}^{2} \ \cos 2\beta \ \} \nonumber \\
&\sim& \frac{24}{16 \pi^{2}} \ln 
\left( \frac{M_{cut}}{M_{W}} \right) \ g_{2}^{6} M_{X}^{2} \ ,
\end{eqnarray}
where $y_{\mu} \equiv m_{\mu}/v$ and 
 $S \equiv m_{Hu}^{2}-m_{Hd}^{2} + {\rm tr} [ m_{Q}^{2}-m_{L}^{2}-2m_{u}^{2}+m_{d}^{2}+m_{e}^{2} ]$.
In deriving (98), we used the relation 
 $M_{1/2}^{a} \sim 4 (g_{4}^{a})^{2} M_{X}$ at the scale $M_{cut}$ 
 shown in (42), and the fact that $M_{1/2}^{a}/(g_{4}^{a})^{2}$ 
 is an RG invariant at the 1-loop level.

Singlet smuon mixes with doublet smuon through the term ${\cal A} m_{\mu}$
 and with singlet stau through the term $c_{\mu \tau} (m_{\mu}/m_{\tau}) M_{X}^{2}$.
Solving these mixings, the mass eigenstate, 
 which is still dominantly composed of singlet smuon,
 has mass estimated as 
\begin{eqnarray}
m^{2}_{{\rm eig}} &\simeq& m_{E2}^{2} 
 + (\frac{m_{\mu}}{m_{\tau}})^{2} M_{X}^{2} \nonumber 
- \frac{ 2 ({\cal A} m_{\mu})^{2} }{ m_{L2}^{2} - m_{E2}^{2} }
 -  \frac{ 2 (c_{\mu \tau} (m_{\mu}/m_{\tau}) M_{X}^{2})^{2} }{ m_{E3}^{2} - m_{E2}^{2} } \nonumber \\
&\simeq& m_{E2}^{2} + \left( \frac{m_{\mu}}{m_{\tau}} \right)^2 M_{X}^{2} 
- \frac{ 2 \left( \mu \ \tan \beta \ m_{\mu} \right)^2 }{ (24/16 \pi^{2}) \ln( M_{cut}/M_{W} ) \ g_{2}^{6} M_{X}^{2} }
 -  \frac{ 2 (c_{\mu \tau} (m_{\mu}/m_{\tau}) M_{X}^{2})^{2} }{ M_{X}^{2} } \nonumber \\
&\simeq& m_{E2}^{2} + (1-2c_{\mu \tau}^{2}) 
\left( \frac{m_{\mu}}{m_{\tau}} \right)^2 M_{X}^{2}
 -  m_{\mu}^{2} \tan^{2} \beta / g_{2}^{6} \ .
\end{eqnarray}
On the other hand, we can neglect the mixing terms 
 between singlet selectron and other sleptons 
 because they are suppressed by the tiny value of $m_{e}/m_{\tau}$.

We now compare the mass of singlet selectron $m_{E1}^{2}$   
 with the mass of the eigenstate, $m_{{\rm eig}}^{2}$.
  Using (97) and (99), we have
\begin{eqnarray}
m_{{\rm eig}}^{2} - m_{E1}^{2} 
&\simeq& - \frac{16}{16 \pi^{2}} 
 \ln \left( \frac{M_{cut}}{M_{W}} \right) M_{X}^{2} \ 
 (y_{\mu})^{2} \tan^{2} \beta 
 +  (1-2c_{\mu \tau}^{2}) \left(\frac{m_{\mu}}{m_{\tau}}\right)^{2} M_{X}^{2}
 -  m_{\mu}^{2} \tan^{2} \beta / g_{2}^{6} \nonumber \\
&\simeq& (1-2c_{\mu \tau}^{2}) 
\left( \frac{m_{\mu}}{m_{\tau}} \right)^2 M_{X}^{2} \ .
\end{eqnarray}
In summary, in the determination of the NLSP mass,
 the effect of the mixing term
 between the singlet smuon and singlet stau 
 dominates over the Yukawa RG contributions and the effects 
 of singlet-doublet mixing terms. 
NLSP is singlet-smuon-like for $c_{\mu \tau} \gtrsim 1/\sqrt{2}$, 
 whereas it is singlet-selectron-like for a relatively small coupling 
 $c_{\mu \tau} \lesssim 1/\sqrt{2}$. 
Since the factor $c_{\mu \tau}$ affects
 $\tau \rightarrow \mu \gamma$ branching ratio 
 through the term $(m_{E}^{2})_{23} = c_{\mu \tau} (m_{\mu}/m_{\tau}) M_{X}^{2}$,
 we have a connection between the value of $Br_{\tau}$ and 
 the flavor of NLSP.

If NLSP is smuon-like, it decays mainly into $\mu$ and gravitino.
However our model predicts that 
 NLSP can contain a considerable amount of stau component
 due to the mixing term $(m_{E}^{2})_{23}$, 
 which can be as large as $(m_{\mu}/m_{\tau}) M_{X}^{2}$. 
Therefore the branching ratio of a flavor-violating NLSP decay
 into $\tau$ and gravitino can be as large as 
\begin{eqnarray}
Br(NLSP \rightarrow \tau \ \psi_{3/2}) 
 \sim (m_{\mu}/m_{\tau})^{2}  \simeq  \frac{1}{300} \ .
\end{eqnarray}
On the other hand, in the context of MFV, the mixing terms 
 of singlet smuon and other sleptons are much smaller 
 as we evaluated in section 5, and such flavor-violating 
 NLSP decays are much suppressed. 
Thus the flavor-violating decay of NLSP provides 
 a distinct signature of our model.
If NLSP is selectron-like, it decays mostly into electron and gravitino.
The model predicts that the mixing terms of singlet selectron 
 and other sleptons are suppressed by $m_{e}$, 
 although they are much larger than in other models with MFV. 
Still, as a distinct signature of our model, 
 we expect to observe a rare NLSP decay into $\tau$ and gravitino
 with the branching ratio as large as 
\begin{eqnarray}
Br(NLSP \rightarrow \tau \ \psi_{3/2}) 
 \sim (m_{e}/m_{\tau})^{2} \ \simeq \ 10^{-7} \ .
\end{eqnarray}
Note that the lifetime of NLSP is estimated as
\begin{eqnarray*}
 t_{NLSP} \simeq 48 \pi 
 \frac{ \vert<F_{\tilde{X}}>\vert^{2}}{m_{NLSP}^5} 
 \simeq  48 \pi \frac{M_{X}^{2} M_{cut}^{2}}{(m_{NLSP})^5} \ ,
\end{eqnarray*} 
which is $\sim 10^{-3}$ sec for $m_{NLSP}=300$ GeV, $M_{X}=500$ GeV and 
 $M_{cut}=10^{13}$ GeV ($\delta_{3} \sim 0.1$). 
Such a long-lived NLSP, once produced at the LHC, 
 decays outside the detector. 
There have been interesting proposals \cite{trap NLSP} 
 for the way to trap charged NLSPs outside the detector. 
Detailed studies of the NLSP decay can allow us 
 not only to measure the gravitino mass and the four-dimensional Planck mass 
 but also to test the flavor-structure of NLSP predicted 
 in our model.

In case (ii), the gravity mediated contributions are negative, 
 so that SU(2) singlet stau is the dominant NLSP component. 
Although this singlet-stau-like NLSP is as usual 
 in SUSY models with gravitino LSP,
 we again expect to observe a rare decay of NLSP 
 as a signature of the model. 
The branching ratio of NLSP decay into $\mu$ and gravitino 
 is predicted to be as large as 
\begin{eqnarray}
Br(NLSP \rightarrow \mu \ \psi_{3/2}) 
 \sim (m_{\mu}/m_{\tau})^{2} \ \simeq \ \frac{1}{300} \ .
\end{eqnarray}

\section{Conclusion}

We have investigated a simple 5D extension of 
 MSSM in RS spacetime, 
 where 5D geometry controls both the SUSY breaking mediation 
 mechanism and the Yukawa coupling hierarchy. 
The Yukawa coupling hierarchy is naturally explained 
 by the localization of matter superfields in the 5D bulk. 
SUSY breaking effects arise from two sources: 
 contact terms between the SUSY breaking sector and 
 the MSSM fields (gravity mediation),
 and RG effects (gaugino mediation). 
The former are flavor-violating and the latter flavor-conserving.
Using the experimental data on fermion masses and mixings, 
 we have determined the 5D disposition of matter superfields and 
 calculated SUSY breaking mass spectra including flavor-violating terms.
We have numerically checked that our framework can give 
 a realistic mass spectrum consistent with 
 all the experimental constraints.

We have estimated the flavor-violating effects 
 induced by RG running in the context of MFV 
 and compared them with those from gravity mediation 
 predicted by the model. 
We have proved that our model provides a different pattern 
 of flavor-violating terms, namely, 
 flavor violation of A-terms and SU(2) singlet soft masses can
 be much larger than the MFV case.

Our model has several distinct predictions.
First, the $\mu \rightarrow e \gamma$ branching ratio 
 is larger than $O(10^{-12})$, regardless of $\tan \beta$ 
 and the seesaw scale, for sparticle masses $\lesssim 2$ TeV. 
This originates from the basic structure of our model, namely, 
 the hierarchy of Yukawa couplings and the gravity mediation
 contributions to A-terms are rooted on the same 5D disposition 
 of the matter superfields.
Hence, using the experimental data 
 on the charged SM fermion masses, CKM matrix 
 and the neutrino oscillation parameters, 
 we can fix the orders of flavor-violating A-terms. 
Second, $\tau \rightarrow \mu \gamma$ branching ratio 
 may not be larger than $\mu \rightarrow e \gamma$ branching ratio.
This is because our model predicts IR-localized singlet stau superfield
 and it gains an additional soft mass through gravity mediation 
 on the IR brane.
Third, since RS geometry warps down the effective cutoff scale,
 which can be characterized by the seesaw scale, 
 gravitino is LSP and the dark matter candidate. 
Forth, NLSP is likely to be either smuon-like or selectron-like 
 because the gravity mediation contribution 
 pushes up the singlet stau mass. 
Furthermore our model predicts flavor-violating NLSP decays 
 with the rates much higher than those expected in the MFV case.

The Yukawa coupling hierarchy is one of the long-standing problems 
 in the Standard Model. 
The 5D MSSM on Randall-Sundrum background offers 
 a solution to this problem from the geometrical point of view. 
In this model, flavor-violating soft SUSY breaking terms have 
 the same geometrical origin as the Yukawa coupling hierarchy. 
Therefore, the flavor structure among sparticles 
 can be a clue to understand the origin of flavors 
 among the SM fermions, even if the origin lies 
 at an energy scale far above the electroweak scale. 
Gravitino is always LSP due to the warped geometry, 
 and NLSP is predicted to be dominantly composed 
 of SU(2) singlet sleptons and long-lived. 
At collider experiments, in this case, supersymmetric events 
 can be fully reconstructed without missing energy, 
 which allows us not only to identify the dominant flavor 
 content of NLSP, but also to measure the rates of 
 flavor-violating decays of sparticles if it is sizable. 
Thus, our framework can be tested in the future.

\section*{Acknowledgments}
The work of NO is supported in part by 
 the DOE Grants, No. DE-FG02-10ER41714. 
The work of TY is supported by a grant of the Japan Society for the Promotion of Science,
 No. 23-3599.
TY would like to thank Department of Physics and Astronomy, 
 University of Alabama for hospitality during his visits.

\newpage
\section*{Appendix: Sample Mass Spectra}

\begin{table}[!h]
\begin{tabular}{|l|r|r|r|r}
\hline
$M_{X}$             & 500   & 600   & 600  \\
$\tan \beta$        & 6     & 5     & 15   \\
$\delta$            & 0.06  & 0.05  & 0.15 \\
\hline \hline
$h^{0}$             & 115.2 & 114.7 & 119.1 \\
$H_{0}$             & 1217  & 1452  & 1423  \\
$H_{A0}$            & 1215  & 1450  & 1420  \\
$H^{\pm}$           & 1218  & 1452  & 1422  \\
\hline 
$\tilde{g}$         & 1845  & 2172 & 2270 \\
$\chi^{0}_{1}$      & 568   & 693  & 647  \\
$\chi^{0}_{2}$      & 814   & 984  & 986  \\
$\chi^{0}_{3}$      & 927   & 1090 & 1133 \\
$\chi^{0}_{4}$      & 964   & 1131 & 1156 \\
$\chi^{\pm}_{1}$    & 815   & 986  & 987  \\
$\chi^{\pm}_{2}$    & 967   & 1135 & 1163 \\
\hline
$\tilde{u}_{L}$     & 1616  & 1893 & 2004 \\
$\tilde{d}_{L}$     & 1618  & 1895 & 2005 \\
$\tilde{u}_{R}$     & 1542  & 1804 & 1912 \\
$\tilde{d}_{R}$     & 1533  & 1793 & 1900 \\
$\tilde{t}_{1}$     & 1288  & 1517 & 1604 \\ 
$\tilde{t}_{2}$     & 1582  & 1848 & 1932 \\ 
$\tilde{b}_{1}$     & 1529  & 1790 & 1878 \\ 
$\tilde{b}_{2}$     & 1552  & 1821 & 1911 \\
\hline
$\tilde{e}_{L}$     & 578   & 685  & 714  \\
$\tilde{e}_{R}$     & 338   & 401  & 414  \\
$\tilde{\nu}_{e}$   & 572   & 681  & 709  \\ 
$\tilde{\tau}_{1}$  & 574   & 683  & 688  \\
$\tilde{\tau}_{2}$  & 611   & 730  & 735  \\
$\tilde{\nu}_{\tau}$& 571   & 680  & 700  \\
\hline \hline
$Br_{\mu} \times 10^{12}$  & 7.6 & 3.9 & 8.2 \\
$Br_{\tau} \times 10^{12}$ & 4.3 & 2.1 & 5.7 \\
\hline
\end{tabular}   
\caption{
Particle mass spectra for three samples for different 
 values of $(M_{X}, \tan \beta,  \delta)$.}
\end{table}

\end{document}